\documentclass[a4paper,11pt]{article}
\pdfoutput=1
\usepackage[normalem]{ulem}
\usepackage[utf8]{inputenc}
\usepackage{cite}
\usepackage[left=2cm,right=2cm,top=2cm,bottom=3cm]{geometry}
\usepackage{xcolor}
\usepackage{float}

\usepackage{tikz-feynman}
\tikzfeynmanset{compat=1.1.0}

\usepackage{epsf}
\usepackage{graphicx}

\usepackage{amsmath}
\usepackage{amssymb}
\usepackage{mathrsfs}
\usepackage{slashed}
\usepackage{mathtools}
\usepackage{bbold}

\usepackage{amsfonts}
\usepackage{xfrac} 
\usepackage{booktabs} 

\usepackage{color}
\definecolor{cred}{RGB}{180,50,40}
\definecolor{purple}{RGB}{180,90,180}
\definecolor{darkgreen}{RGB}{0, 100, 0}

\usepackage[
colorlinks=true
,urlcolor=black
,anchorcolor=black
,citecolor=black
,filecolor=black
,linkcolor=black
,menucolor=black
,linktocpage=true
,pdfproducer=medialab
,pdfa=true
]{hyperref}

\usepackage{braket}

\begin{document}

\allowdisplaybreaks
\setcounter{footnote}{0}
\begin{center}
\vspace*{1mm}
\vspace{1cm}
{\Large\bf
Anomalies in $\pmb{^8}$Be nuclear transitions and 
$\pmb{(g-2)_{e,\mu}}$: \\ towards a
minimal combined explanation} 

\vspace*{0.8cm}

{\bf C.~Hati $^{a}$, J.~Kriewald $^{b}$, J. Orloff $^{b}$ and
  A.~M.~Teixeira $^{b}$} 

\vspace*{.5cm}
$^{a}$ Physik Department T70, Technische Universit\"at M\"unchen,\\
James-Franck-Stra{\ss}e 1, D-85748 Garching, Germany

\vspace*{.2cm}
$^{b}$Laboratoire de Physique de Clermont (UMR 6533), CNRS/IN2P3,\\
Univ. Clermont Auvergne, 4 Av. Blaise Pascal, F-63178 Aubi\`ere Cedex,
France
\end{center}

\vspace*{2mm}
\begin{abstract}
\noindent
Motivated by a simultaneous explanation of the apparent
discrepancies in the light charged lepton anomalous magnetic dipole
moments, and the anomalous 
internal pair creation in $^8$Be nuclear transitions, we
explore a simple New Physics model, based on an extension of the
Standard Model gauge group by a $U(1)_{B-L}$. The model further
includes heavy vector-like 
fermion fields, as well as an extra scalar responsible for the low-scale
breaking of $U(1)_{B-L}$, which gives rise to a light 
$Z^\prime$ boson. 
The new fields and currents allow to explain the
anomalous internal pair creation in $^8$Be while being consistent with various
experimental constraints. Interestingly, we find that 
the contributions of the $Z^\prime$ and the new
$U(1)_{B-L}$-breaking scalar can also successfully account for both 
$(g-2)_{e,\mu}$ anomalies; the strong phenomenological constraints on
the model's parameter space ultimately render the 
combined explanation of $(g-2)_e$ and the anomalous internal pair
creation in $^8$Be particularly predictive. The underlying idea of this minimal 
``prototype model'' can be readily incorporated into other 
protophobic $U(1)$ extensions of the Standard Model.

\end{abstract}
\section{Introduction}\label{sec:intro}

The possible existence of new interactions, in addition to those
associated with the Standard Model (SM) gauge group, has been a
longstanding source of interest, both for particle and astroparticle physics.
Numerous experimental searches have been dedicated to look for
theoretically well-motivated light mediators, such as 
axions (spin-zero), dark photons (spin-1) or light $Z^\prime$ (spin-1)~\cite{Ni:1999di,Heckel:2006ww,Baessler:2006vm,Hammond:2007jm,Heckel:2008hw,Vasilakis:2008yn,Serebrov:2009ej,Ignatovich:2009zz,Serebrov:2009pa,Karshenboim:2010cg,Karshenboim:2010cj,Petukhov:2010dn,Karshenboim:2011dx,Hoedl:2011zz,Raffelt:2012sp,Yan:2012wk,Tullney:2013wqa,Chu:2012cf,Bulatowicz:2013hf,Mantry:2014zsa,Hunter:2013gba,Salumbides:2013dua,Leslie:2014mua,Arvanitaki:2014dfa,Stadnik:2014xja,Afach:2014bir,Terrano:2015sna,Leefer:2016xfu,Ficek:2016qwp,Ji:2016tmv,Crescini:2017uxs,Dzuba:2017puc,Delaunay:2017dku,Stadnik:2017hpa,Rong:2017wzk,Safronova:2017xyt,Ficek:2018mck,Rong:2018yos,Dzuba:2018anu,Kim:2017yen
}. 

Several distinct probes have been used to look for the presence of the
new mediators. 
Nuclear transitions are among the most interesting and promising
laboratories to search for relatively light new physics states. 
A few years ago, the Atomki Collaboration 
reported their results~\cite{Krasznahorkay:2015iga} 
on the measurement of the angular
correlation of electron-positron internal pair creation (IPC) for two
nuclear transitions of Beryllium atoms
($^8$Be), with a significant excess being observed  at large angles
for one of them.
The magnetic dipole ($M1$) transitions under study
concerned the decays of the excited isotriplet and isosinglet states, 
respectively denoted $^8$Be$^{*'}$ and $^8$Be$^{*}$, into the
fundamental state ($^8$Be$^{0}$). The transitions are summarised
below, together with the associated energies:
\begin{align}
^8\text{Be}^{*'} (j^\pi = 1^+,T=1)\, 
\rightarrow \, 
^8\text{Be}^{0} (j^\pi = 0^+,T=0)\,, \quad 
E\,=\, 17.64\text{ MeV}\,; \nonumber \\
^8\text{Be}^{*} (j^\pi = 1^+,T=0)
\,\rightarrow\, 
^8\text{Be}^{0} (j^\pi = 0^+,T=0)\,, \quad 
E\,=\, 18.15\text{ MeV}\,,
\end{align}
in which $j^\pi$ and $T$ correspond
to the spin-parity and isospin of the nuclear states, respectively. A significant enhancement of the IPC was observed at large angles in the
angular correlation of the 18.15~MeV transition; it was subsequently
pointed out that such an anomalous result could be potentially interpreted
as the creation and decay of an unknown intermediate light particle with
mass $m_{X}$=16.70$\pm0.35 $(stat)$\pm 0.5 $(sys)~MeV~\cite{Krasznahorkay:2015iga} .  

Recently, a re-investigation of the $^8$Be anomaly with an
improved set up corroborated the earlier 
results for the 18.15~MeV transition~\cite{Krasznahorkay:2017qfd,Krasznahorkay:2017bwh,Krasznahorkay:2018snd,Krasznahorkay:2019lgi};
moreover, it allowed constraining the mass of the hypothetical mediator
to $m_X = 17.01(16)$~MeV and its branching ratio 
(normalised to the $\gamma$-decay) to $\Gamma_X/\Gamma_\gamma = 6(1)\times 10^{-6}$. The $e^+e^-$ pair correlation in the 17.64~MeV
transition of $^8$Be was also revisited, and again no significant
anomalies were found~\cite{Krasznahorkay:2017gwn,Gulyas:2015mia}. A
combined interpretation of the data of $^8\text{Be}^*$ decays 
(only one set exhibiting an anomalous behaviour)
in terms of a new light particle, in association with
the possibility of mixing between the two different excited $^8$Be
isospin states ($^8\text{Be}^{*'}$ and $^8\text{Be}^{*}$)
might suggest a larger preferred mass
for the new mediator; this would lead to a large phase space suppression,
therefore potentially explaining the null results for
$^8\text{Be}^{*'}$ decay. In turn, it can further entail 
significant changes in the preferred quark (nucleon)
couplings to the new particle mediating the anomalous IPC, 
corresponding to significantly smaller normalised branching fractions 
than those of the preferred fit of~\cite{Krasznahorkay:2019lyl}. 

Further anomalies in nuclear transitions have been observed, in
particular concerning the 21.01~MeV  $0^-\rightarrow 0^+$  transition
of $^4$He~\cite{Krasznahorkay:2019lyl,Firak:2020eil}, resulting in 
another anomalous IPC corresponding to the angular correlation of
electron-positron pairs at 115$^\circ$, with 7.2$\sigma$ significance. 
The result can also be potentially interpreted as the creation and
subsequent decay of a light particle: the corresponding mass and
width, $m_{X}$=16.98$\pm0.16\text{(stat)}\pm0.20 (\text{syst})$~MeV, 
and $\Gamma_{X}$= $3.9\times
10^{-5}$~eV, lie in a range similar to that suggested by the anomalous 
$^8\mathrm{Be}$ transition.

If the anomalous IPC observations are to be interpreted as being
mediated by a light new state\footnote{In Ref.~\cite{Zhang:2017zap}, 
the possibility of explaining the anomaly within
nuclear physics was explored; however, the required form factors
were found to be unrealistic for a nucleus like $^8$Be.}, 
the latter can a priori be a scalar,
pseudoscalar, vector, axial vector, or even a spin-2 particle,
provided it decays into electron-positron pairs. 
For a parity conserving scenario,
the hypothesis of an intermediate scalar boson has already been
dismissed~\cite{Feng:2016jff}, due conservation of angular momentum in
the $1^+ \rightarrow 0^+$ $^8$Be transition. 
Having a pseudoscalar mediator has been also severely constrained (and disfavoured) by experiments -- for an axion-like particle $a$ with a mass of $m_a\approx 17~\text{MeV}$ and an interaction term 
$g_{a} a F^{\mu\nu}\tilde{F}_{\mu\nu}$, all couplings in the range 
$ 1 /(10^{18}~\text{GeV}) < g_{a} < 1/ (10~\text{GeV})$ are
excluded~\cite{Hewett:2012ns,Dobrich:2015jyk} (although this can 
be partially circumvented in the presence of additional non-photonic
couplings~\cite{Ellwanger:2016wfe}). 
A potential first explanation of the anomalous IPC in $^8$Be 
in the context of simple $U(1)$ extensions of the
SM was discussed in~\cite{Feng:2016jff,Feng:2016ysn}, relying on the 
exchange of a 16.7~MeV, $j^\pi$ = 1$^+$
vector gauge boson. 
In~\cite{Ellwanger:2016wfe} the possibility of a light pseudoscalar
particle with $j^\pi = 0^-$ was examined, while a potential
explanation based on an axial vector particle (including an ab-initio
computation for the relevant form factors) was carried
in~\cite{Kozaczuk:2016nma}. 
Further ideas were put forward and discussed (see, for 
instance,~\cite{Gu:2016ege,Seto:2016pks,DelleRose:2017xil,DelleRose:2018eic,BORDES:2019wcp,Nam:2019osu,Pulice:2019xel,Wong:2020hjc,Tursunov:2020wfy,Kirpichnikov:2020tcf,Koch:2020ouk,Jentschura:2020zlr}). 

The favoured scenario of a new light vector boson is nevertheless
heavily challenged by numerous experimental constraints: 
the dark photon hypothesis is strongly disfavoured in view of negative
searches for associate production in rare light meson decays 
(e.g., $\pi^0 \rightarrow \gamma A'$ at NA48/2 which, for a dark
photon mass $\mathcal{O}(17~\text{MeV})$, constrains its
couplings to be strictly ``protophobic'', in stark contrast with the
requirements to explain the anomalous IPC in $^8$Be); the
generalisation towards a protophobic vector boson arising from 
a gauged $U(1)$ extension of the SM (potentially with
both vector and axial couplings to the SM fields) is also subject 
to stringent constraints from the measurements of atomic parity
violation in Caesium (Cs) and neutrino-electron scattering (as well as
non-observation of coherent neutrino–nucleus scattering), which force 
the leptonic couplings of the gauge boson to be too small to account
for the anomalous IPC in $^8$Be. Interestingly this problem can be
circumvented in the presence of additional vector-like leptons as
noted in~\cite{Feng:2016ysn} - an idea we will pursue and explore in
our work.

Extensions of the SM which include light new physics states coupled to
the standard charged leptons are a priori expected to have 
implications for precision tests of leptonic observables, and even have the potential to address (solving, or at least rendering less severe)
other anomalies, as is the case of those concerning the anomalous magnetic
moment of light charged leptons, usually expressed
in terms of $a_\ell \equiv (g-2)_{\ell}/2$ ($\ell = e, \mu$).  
As of today (while expecting new data from the new $(g-2)_\mu$
experiment at FNAL~\cite{Grange:2015fou}),    
a long standing tension persists between the muon's 
measured value~\cite{Bennett:2006fi,Mohr:2015ccw}
\begin{equation}\label{eq:amu:exp}
a_\mu^\text{exp}\,=\,116,\!592,\!089(63)\times 10^{-11}\,,
\end{equation}
and the SM prediction with improved hadronic vacuum
polarisation~\cite{Chakraborty:2016mwy,Jegerlehner:2017lbd,DellaMorte:2017dyu,Davier:2017zfy,Borsanyi:2017zdw,Blum:2018mom,Keshavarzi:2018mgv,
Colangelo:2018mtw,Davier:2019can}, light-by-light
scattering~\cite{Colangelo:2015ama,Green:2015sra,Gerardin:2016cqj,Blum:2016lnc,Colangelo:2017qdm,Colangelo:2017fiz,Blum:2017cer,Hoferichter:2018dmo,
Hoferichter:2018kwz}, and higher-order hadronic
corrections~\cite{Kurz:2014wya,Colangelo:2014qya}, typically leading to
deviations of about $3\sigma$ from the experimental 
result~\footnote{Recent results of a lattice study of leading
  order hadronic vacuum polarisation suggested that the 
  discrepancy could be significantly reduced~\cite{Borsanyi:2020mff}; 
  however, in~\cite{Crivellin:2020zul} it was pointed out that such hadronic
  vacuum polarisation contributions could potentially lead to conflicts
  with electroweak fits, inducing tensions in other relevant
  observables (hitherto in good agreement with the SM).}, 
\begin{equation}
\label{Delta_amu}
\Delta a_\mu\,=\,a_\mu^\text{exp} - a_\mu^\text{SM}
\,\sim \,2.7 (0.7)\times 10^{-9}\,.
\end{equation}
Interestingly, a precise measurement of $\alpha_e$ 
using $\mathrm{Cs}$ atoms~\cite{Parker:2018vye,Yu:2019gdh} has recently given rise to yet another discrepancy, this time
concerning the electron's anomalous magnetic moment. The 
experimental measurement of the electron anomalous magnetic moment
$a_e$~\cite{Hanneke:2008tm} 
\begin{equation}
a_e^\text{exp}\, =\, 1,\!159,\!652,\!180.73(28)\times 10^{-12}\,
\end{equation}
currently exhibits a $2.5\sigma$ deviation from 
the SM prediction, 
\begin{equation}
\label{Delta_ae}
\Delta a_e\,=\,a_e^\text{exp} - a_e^\text{SM}\,\sim\, 
-0.88 (0.36)\times 10^{-12}\,.
\end{equation}
Notice that in addition to being by themselves deviations from the SM
expectations, the joint behaviour of $\Delta a_{e,\mu}$ is also
puzzling: not only is 
the sign of $\Delta a_e$ opposite to that of $\Delta a_\mu$, but 
the ratio $\Delta a_\mu/\Delta a_e$ does not follow the na\"ive
quadratic behaviour $m_\mu^2/m_e^2$ for the magnetic dipole operator 
(where the necessary chirality flip appears through a mass insertion
of the SM lepton)~\cite{Giudice:2012ms}. 
It thus becomes particularly challenging to explain both anomalies 
simultaneously within a minimal flavour violation (MFV) hypothesis,
or even within minimal SM extensions via a single new 
particle coupling to charged
leptons~\cite{Davoudiasl:2018fbb,Kahn:2016vjr,Crivellin:2018qmi}. 
The observed pattern for $\Delta a_e$ and $\Delta a_\mu$ further
suggests the underlying presence of New Physics (NP) potentially
violating the universality of
lepton flavours. Several attempts have been recently conducted to
simultaneously explain the tensions in both $\Delta a_{e,\mu}$
(see for example~\cite{Crivellin:2018qmi,Liu:2018xkx,Dutta:2018fge,Han:2018znu,Endo:2019bcj,Kawamura:2019rth,Abdullah:2019ofw,Badziak:2019gaf,CarcamoHernandez:2019ydc,Hiller:2019mou,Cornella:2019uxs,CarcamoHernandez:2020pxw,Haba:2020gkr,Bigaran:2020jil,Jana:2020pxx,Calibbi:2020emz,Chen:2020jvl,Yang:2020bmh}): in particular, certain 
scenarios have explored a chiral enhancement, 
due to heavy vector-like leptons in the
one-loop dipole operator, which can potentially lead to sizeable
contributions for the leptonic magnetic moments; 
however, this can open the door to charged lepton flavour violating
interactions (new physics fields with non-trivial
couplings to both muons and electrons can potentially lead to sizeable
rates for $\mu\rightarrow e\gamma, \mu\rightarrow 3e$ and $\mu-e$
conversion), already in conflict with current data~\cite{Tanabashi:2018oca}.
Controlled couplings of
electrons and muons to beyond the Standard Model (BSM) fields in the loop (subject to
``generation-wise'' mixing between SM and heavy vector-like fields)
can be achieved, 
and this further allows to evade the potentially 
very stringent constraints from charged lepton flavour
violating (cLFV) $\mu-e$ transitions.

\bigskip
In this work, we explore a simple New Physics model, based on an 
extended gauge group 
$SU(3)\times SU(2)_L \times U(1)_Y \times U(1)_{B-L}$, with the SM
particle content extended by heavy vector-like 
fermion fields, in addition to the light 
$Z^\prime$ associated with a low-scale breaking of $U(1)_{B-L}$ by an
extra scalar field. This ``prototype model'' offers a minimal
scenario to successfully explain the anomalous 
internal pair creation in $^8$Be 
while being consistent with various experimental bounds. 
However, the couplings of the light $Z^\prime$ to fermions are strongly
constrained by experimental searches: the measurement of the atomic parity
violation in Caesium proves to be one of the most stringent constraints in
what concerns couplings to the electrons. Likewise, neutrino-electron
scattering and the non-observation of 
coherent neutrino–nucleus scattering impose equally stringent constraints on $Z^\prime$-neutrino couplings (the tightest bounds being due to the TEXONO~\cite{Deniz:2009mu} and CHARM-II~\cite{Vilain:1993kd} experiments). 

Our findings reveal that the interplay of the (one-loop)
contributions of the $Z^\prime$ and the $U(1)_{B-L}$ breaking Higgs
scalar can further saturate the discrepancies in both 
$(g-2)_{e,\mu}$ anomalies. In particular, 
a cancellation between the new contributions 
is crucial to reproduce the observed pattern of
opposite signs of $\Delta a_e$ and $\Delta a_\mu$.
In view of the extensive limits on the $Z^\prime$ couplings,
arising from experimental searches, and which are 
further constrained by the requirements to explain  
the anomalous IPC in $^8$Be atoms, 
a combined explanation of the different anomalies renders the model
ultimately predictive in what concerns the electron 
$(g-2)$. 
We emphasise that even though we have considered a particular
$U(1)_{B-L}$ extension here - a minimal working
``prototype model'' - the general idea can be straightforwardly 
adopted and incorporated into other possible protophobic $U(1)$ extensions of the SM.

The model is described in~Section~\ref{sec:model}, 
in which we detail the couplings of the exotic states to SM fields,
and their impact for the new neutral current interactions.
After a brief description of the new contributions to charged lepton
anomalous magnetic moments (in a generic way) in 
Section~\ref{sec:g-2}, Section~\ref{sec:IPCcon}
is devoted to discussing
how a the light $Z^\prime$ can successfully explain the several
reported results on the anomalous IPC in $^8$Be atoms, including a 
discussion of potentially relevant isospin-breaking effects.
We revisit the available experimental constraints in 
Section~\ref{sec:phenocon},
and subsequently investigate how these impact the model's parameter
space in Section~\ref{sec:combined:explanation}, in particular 
the viable regimes allowing for a combined explanation of 
$^8$Be anomaly, as well as the tensions in $(g-2)_{e,\mu}$. 
A summary of the key points and further discussion is given in the 
Conclusions. 


\section{A light vector boson from a $\pmb{U(1)_{B-L}}$: the model}
\label{sec:model}
We consider a minimal gauge extension of the SM gauge group,
$SU(3)\times SU(2)_L \times U(1)_Y \times U(1)_{B-L}$.  
Such an extension with a locally gauged $U(1)_{B-L}$ gives rise
to new gauge and gauge-gravitational anomalies in the theory, which
need to be cancelled.
In particular, the gauged $U(1)_{B-L}$ gives rise to the triangular
gauge anomalies -  
$\mathcal{A}\left[U(1)_{B-L}\left(SU(2)_{L}\right)^2\right]$,
$\mathcal{A}\left[\left(U(1)_{B-L}\right)^3\right]$,
$\mathcal{A}\left[U(1)_{B-L} \left(  U(1)_{Y}\right)^2\right]$,  and
$\mathcal{A}\left[\mbox{Gravity}^2 \times U(1)_{B-L}\right]$. 
While the first two automatically vanish for the SM field content,
the other two require a (positive) contribution from additional fields. 
One of the most conventional and economical ways to achieve this
relies on the introduction of a SM singlet neutral fermion ($N_R$),
with a charge $B-L=-1$, for each fermion generation. 
In the present model, 
the $U(1)_{B-L}$ is broken at a low scale by a SM singlet scalar, $h_X$,
which acquires a vacuum expectation value (VEV) $v_X$,
responsible for a light vector boson, with a mass $m_{Z^\prime} 
\sim \mathcal{O}(17~\text{MeV})$.
A successful explanation of the $^8$Be anomaly
through a light $Z^\prime$~\cite{Feng:2016ysn} further requires the
presence of additional fields ($L$ and $E$) to ensure phenomenological
viability in view of the constraints from various
experiments\footnote{In particular, constraints arising from
  neutrino-electron scattering experiments and atomic parity violation
  require the addition of this exotic particle content as
  discussed in more detail in Section~\ref{sec:phenocon}.};
thus, the model also
includes three generations of isodoublet and isosinglet vector-like
leptons.
The field content of the model and the transformation properties under
the extended gauge group $SU(3)\times SU(2)_L \times U(1)_Y \times U(1)_{B-L}$
are summarised in Table~\ref{tab:fields}. 
\begin{table}[ht!]
\begin{center}
  \begin{tabular}{|c|c|c|c|c|}
  \hline
      Field & $SU(3)_c$ & $SU(2)_L$ & $U(1)_Y$ & $U(1)_{B-L}$\\
  \hline
  \hline
  $Q = \left(u_L, \, d_L\right)^T$ & $ \mathbf{3} $ & $ \mathbf{2} $ &
  $ \frac{1}{6} $ & $ \frac{1}{3} $\\ 
  $ \ell = \left(\nu_L, \, e_L\right)^T $ & $ \mathbf{1} $ & $
  \mathbf{2} $ & $ -\frac{1}{2} $ & $ -1 $\\ 
  $ u_R $ & $ \mathbf{3} $ & $ \mathbf{1} $ & $ \frac{2}{3} $ & $
  \frac{1}{3} $\\ 
  $ d_R $ & $ \mathbf{3} $ & $ \mathbf{1} $ & $ -\frac{1}{3} $ & $
  \frac{1}{3} $\\ 
   $ e_R $ & $ \mathbf{1} $ & $ \mathbf{1} $ & $ -1 $ & $ -1 $\\
  \hline
  $ h_\text{SM} $ & $ \mathbf{1} $ & $ \mathbf{2} $ & $ \frac{1}{2} $ & $ 0 $ \\
  \hline
  \hline
  $ N_R$ & $ \mathbf{1} $ & $ \mathbf{1} $ & $ 0 $ & $ -1 $\\
  $ L_{L,R} = \left( L_{L,R}^0 , \, L_{L,R}^-\right)^T $ & $
  \mathbf{1} $ & $ \mathbf{2} $ & $ -\frac{1}{2} $ & $ 1 $ \\ 
  $ E_{L,R} $ & $ \mathbf{1} $ & $ \mathbf{1} $ & $ -1 $ & $ 1 $ \\
  \hline
  $ h_{X} $ & $ \mathbf{1} $ & $ \mathbf{1} $ & $ 0 $ & $ 2 $\\
  \hline
  \end{tabular}
  \end{center}
\caption{Field content of the model and
  transformation properties under the gauge group $SU(3)\times SU(2)_L
  \times U(1)_Y \times U(1)_{B-L}$.} 
  \label{tab:fields}
\end{table}
%


\subsection{Gauge sector}\label{sec:mixing}
In the unbroken phase, the relevant part of the kinetic Lagrangian,
including mixing~\footnote{We recall that
  kinetic mixing always appears at least at the one-loop level
  in models with fermions which are charged under
  both $U(1)$s. Here we parametrise these corrections through an effective
  coefficient, $\epsilon_k$.} between the hypercharge boson $B$ and
the $U(1)_{B-L}$ boson $B^\prime$, is given by 
\begin{equation}
  \mathcal L^\text{gauge}_\text{kin.} \supseteq
  -\frac{1}{4}\tilde{F}_{\mu\nu} \tilde{F}^{\mu\nu}
  - \frac{1}{4} \tilde{F^\prime}_{\mu\nu}\tilde{F^\prime}^{\mu\nu}
  + \frac{\epsilon_k}{2}\,\tilde{F}_{\mu\nu}\tilde{F^\prime}^{\mu\nu} +
  \sum_{f} i \bar f \,\tilde{\slashed{D}}\, f\,.
\end{equation}
In the above, $\tilde{F}_{\mu\nu}$ and $\tilde{F^\prime}_{\mu\nu}$
correspond to the field strengths of the (kinetically mixed)
hypercharge boson $\tilde B$ and the $U(1)_{B-L}$ boson $\tilde
B^\prime$;
$\epsilon_k$ denotes the kinetic mixing parameter.
The relevant part of the gauge covariant derivative is given by
\begin{equation}
  \tilde{D}_\mu \,=  \,\partial_\mu + \dots + i \,g^\prime \, Y_f \,
  \tilde{B}_\mu
  + i \,g_{B-L}  \,Q_f^{B-L} \, \tilde{B}^\prime_\mu\,, 
\end{equation}
with the hypercharge and $B-L$ charge written as
$Y_f = Q_f-T_{3\,f}$ and $Q^{B-L}_f$, 
respectively;
the corresponding gauge couplings are denoted by $g^\prime$ and $g_{B-L}$.
The gauge kinetic terms can be cast in matrix form as
\begin{equation}
	-\frac{1}{4} \tilde{F}_{\mu\nu}\tilde{F}^{\mu\nu} -\frac{1}{4}
        \tilde{F^\prime}_{\mu\nu}\tilde{F^\prime}^{\mu\nu} + \frac{\epsilon_k}{2}
        \tilde{F}_{\mu\nu}\tilde{F^\prime}^{\mu\nu} =
        -\frac{1}{4}\begin{pmatrix} \tilde{F}_{\mu\nu} &
          \tilde{F^\prime}_{\mu\nu}\end{pmatrix} 
	\begin{pmatrix} 1 & -\epsilon_k\\
          -\epsilon_k & 1\end{pmatrix}
	  \begin{pmatrix}\tilde{F}^{\mu\nu} \\
            \tilde{F^\prime}^{\mu\nu}\end{pmatrix}\,,
\end{equation}
which can then be brought to the diagonal canonical form
\begin{equation}
  \mathcal{L}^\text{gauge}_\text{kin.} =
  -\frac{1}{4} F_{\mu\nu}F^{\mu\nu} -\frac{1}{4} F^\prime_{\mu\nu}F^{\prime\mu\nu}
  + \sum_f i \bar f \,{\slashed{D}} \,f
\end{equation}
by a linear transformation of the fields,
\begin{equation}
  \tilde{B}_\mu = B_\mu + \frac{\epsilon_k}{\sqrt{1 -
      \epsilon_k^2}}B^\prime_\mu\:\text,\quad\quad
  \tilde{B^\prime}_\mu = \frac{1}{\sqrt{1-\epsilon_k^2}} B^\prime_\mu
  \, . 
\end{equation}
This transformation is obtained by a Cholesky decomposition, allowing
the resulting triangular matrices to be absorbed into a redefinition
of the gauge fields. 
The neutral part of the gauge covariant derivative can then be written as
\begin{equation}
  D_\mu \,=  \,\partial_\mu + \dots + i  \,g \, T_{3\,f}  \,W_{3\,\mu}
  + i  \,g^\prime  \,Y_f \,B_\mu + i\left(\varepsilon^\prime  
\,g^\prime  \,Y_f
  + \varepsilon^\prime_{B-L} \,  Q^{B-L}_f\right) B^\prime_\mu \,, 
  \label{eqn:covdiv}
\end{equation}
in which we have introduced the following coupling strengths
\begin{equation}\label{eq:def:epsilonprime}
  \varepsilon^\prime_{B-L} = \frac{g_{B-L}}{\sqrt{1 - \epsilon_k^2}} \,,
  \quad\quad 
\varepsilon^\prime = \frac{\epsilon_k}{\sqrt{1-\epsilon_k^2}}\:\text.
\end{equation}
Note that due to the above transformation, the mixing now appears
in the couplings of the physical fields.
In the broken phase 
(following electroweak symmetry breaking, and the
subsequent $U(1)_{B-L}$ breaking), the Lagrangian includes the following mass terms 
\begin{equation}
  \mathcal{L}^\text{gauge}_\text{mass}
  \supseteq (D_\mu\braket{h_\text{SM}})^\dagger \,
  (D^\mu\braket{h_\text{SM}}) + (D_\mu\braket{h_{X}})^\dagger\,
  (D^\mu\braket{h_{X}})\,,
\end{equation}
with the covariant derivative $D_\mu$ defined in Eq.~(\ref{eqn:covdiv}).
The resulting mass matrix, in which the neutral bosons mix amongst
themselves, can be diagonalised, leading to the following relations
between physical and interaction gauge boson states
 \begin{equation}
  \begin{pmatrix}A^\mu\\ Z^\mu\\Z^{\prime\,\mu}\end{pmatrix}
  = \begin{pmatrix}
  \cos\theta_w & \sin\theta_w & 0\\
  - \sin\theta_w \cos\theta^\prime & \cos\theta_w\cos\theta^\prime &
  \sin\theta^\prime\\ 
  \sin\theta_w\sin\theta^\prime & -\cos\theta_w\sin\theta^\prime &
  \cos\theta^\prime 
  \end{pmatrix}
  \begin{pmatrix}B^\mu\\
    W_3^\mu\\B^{\prime\,\mu}\end{pmatrix}\text,
 \end{equation}
with the mixing angle $\theta^\prime$ defined as
\begin{equation}
  \tan2\theta^\prime \,=\,
  \frac{2\,\varepsilon^\prime \,g^\prime\, \sqrt{g^2 + {g^\prime}^2}}{
    {\varepsilon^\prime}^2\,{g^\prime}^2 + 4\, m_{B^\prime}^2/v^2 - g^2
    - {g^\prime}^2} \, ,
\end{equation}
in which $m_{B^\prime}^2 = 4 \,{\varepsilon^\prime}_{B-L}^2 \,v_X^2$ is the mass
term for the $B^\prime$-boson induced by $v_X$ (the VEV of the scalar singlet $h_X$ responsible for $U(1)_{B-L}$
breaking), and $\theta_w$ is the standard weak
mixing angle. The mass eigenvalues of the neutral vector bosons are given by 
\begin{equation}
M_{A} = 0\,, \quad \quad
M_{Z,\,Z^\prime} = \frac{g}{\cos\theta_w}
\frac{v}{2}\left[\frac{1}{2}\left(\frac{{\varepsilon^\prime}^2 +
    {4\,m_{B^\prime}^2}/{v^2}}{g^2 + {g^\prime}^2} + 1\right) \mp
  \frac{g^\prime\,\cos\theta_w\,
  \varepsilon^\prime}{g\,\sin2\theta^\prime}
  \right]^{\frac{1}{2}}\text.  
\end{equation}
In the limit of small $\varepsilon^\prime$ (corresponding to small
kinetic mixing, cf. Eq.~(\ref{eq:def:epsilonprime}),
one finds the following approximate expressions for
the mixing angle and the masses of the $Z$ and $Z^\prime$ bosons,
\begin{equation}\label{eq:mix2}
  M_Z^2 \simeq \frac{g^2 + {g^\prime}^2}{4} \,v^2\,,\quad M_{Z^\prime}^2\simeq
  m_{B^\prime}^2\,,\quad \tan2\theta^\prime \simeq -2
  \varepsilon^\prime \,\sin\theta_w\,.
\end{equation}
The relevant terms of the gauge covariant derivative
can now be expressed as~\footnote{Corrections in the $Z$ coupling due
  to mixing with the $Z^\prime$ only appear at order
  ${\varepsilon^\prime}^2\:\text{or}\:\varepsilon^\prime\varepsilon^\prime_{B-L}$
  and will henceforth be neglected.} 
\begin{equation}\label{eq:mix3}
  D_\mu \,\simeq \,\partial_\mu + \dots + i \frac{g}{\cos\theta_w}\,
  (T_{3\,f} - \sin^2\theta_W \,Q_f)\,Z_\mu + ie\, Q_f \,A_\mu +
  ie\,(\varepsilon\,
  Q_f + \varepsilon_{B-L}\,Q_f^{B-L})\,Z^\prime_\mu\, , 
\end{equation}
in which the kinetic mixing parameter and the $B-L$ gauge
coupling have been redefined as 
\begin{equation}\label{eq:epsilon:redefine}
\varepsilon \,= \, {\varepsilon^\prime
  \cos\theta_w}\,, \quad \quad
\varepsilon_{B-L} \,=\,
{\varepsilon^\prime_{B-L}}/{e}\,.
\end{equation}

\subsection{Lepton sector: masses and mixings}
Fermion masses (both for SM leptons and the additional vector-like
leptons) arise from the following generic terms in the Lagrangian
\begin{eqnarray}
  \mathcal{L}^\text{lepton}_{\text{mass}} &=&
  -y_\ell^{ij} \,h_\text{SM} \,\bar\ell_L^i \,e_R^j
  + y_\nu^{ij} \,\tilde{h}_\text{SM} \,\bar\ell_L^i \,N_R^j -\frac{1}{2}\,y_M^{ij}
  \,h_{X} \,\bar N_R^{i\,c} \,N_R^j -
  \lambda_L^{ij} \,h_{X}\, \bar\ell_L^i\,L_R^j -
  \lambda_E^{ij} \,h_{X}\,\bar E_L^{i} \,e_R^j\nonumber\\ 
  &\phantom{k}& - M_L^{ij} \,\bar L_L^i \,L_R^j - M_E^{ij} \,\bar E_L^i\,
  E_R^j - h^{ij}\, h_\text{SM} \,\bar L_L^i \,E_R^j + k^{ij} \,
  \tilde{h}_\text{SM} \,\bar
  E_L^i \,L_R^j + \mathrm{H.c.}\,,
  \label{eqn:yuk}
\end{eqnarray}
in which $y$, $\lambda$, $k$ and $h$ denote Yukawa-like interactions
involving the SM leptons, heavy right-handed neutrinos and the vector-like 
neutral and charged leptons; as mentioned in the beginning of this
section, and in addition to the three SM generations of neutral and
charged leptons (i.e., 3 flavours), the model includes three
generations of isodoublet and isosinglet vector-like leptons.
In Eq.~(\ref{eqn:yuk}), each coupling or mass term thus runs over
$i,j=1...3$, i.e. $i$ and $j$ denote the three generations intrinsic
to each lepton species.

As emphasised in Ref.~\cite{Crivellin:2018qmi}, 
intergenerational couplings between the SM charged leptons and
the vector-like fermions should be very small, in order to avoid
the otherwise unacceptably large rates
for cLFV processes,
as for instance $\mu\to e\gamma$. In what follows, and to further
circumvent excessive
flavour changing neutral current (FCNC) interactions mediated by the
light $Z^\prime$,
we assume the couplings $h$, $k$, $\lambda_L$ and $\lambda_E$, 
as well as the masses $M_{L,E}$, to be diagonal,
implying that the SM fields (neutrinos and charged leptons)
of a given generation can only mix with vector-like fields
of the same generation. 

After electroweak and $U(1)_{B-L}$ breaking,
the mass matrices for the charged leptons and neutrinos can be cast
for simplicity in a ``chiral basis'' spanned, for each generation, by
the following vectors: $(e_L,L_L^-,E_L)^T$, $(e_R,L_R^-,E_R)^T$ and
$(\nu, N^c,L^{0},L^{0c})_L^T$. 
The charged lepton mass matrix is thus given by
\begin{equation}
  \mathcal L^{\ell}_{\text{mass}} =\begin{pmatrix}\bar e_L& \bar
  L_L^-& \bar E_L\end{pmatrix} 
  \cdot M_\ell \cdot
  \begin{pmatrix}e_R\\L_R^-\\E_R\end{pmatrix}
  = \begin{pmatrix}\bar e_L& \bar L_L^-& \bar E_L\end{pmatrix}
  \begin{pmatrix}
    y \frac{v}{\sqrt{2}} & \lambda_L \frac{v_{X}}{\sqrt{2}} & 0 \\
    0 & M_L & h \frac{v}{\sqrt{2}}\\
    \lambda_E \frac{v_{X}}{\sqrt{2}} & k \frac{v}{\sqrt{2}} & M_E\\
  \end{pmatrix}
  \begin{pmatrix}e_R\\L_R^-\\E_R\end{pmatrix}\,\text,
\end{equation}
in which every entry should be understood as a $3\times3$ block (in
generation space).
The full charged lepton mass matrix can be (block-) diagonalised by a
bi-unitary rotation 
\begin{equation}\label{eq:mix4}
  M_\ell^\text{diag} = U_L^\dagger \,M_\ell\, U_R\,,
\end{equation}
where the rotation matrices $U_{L,R}$ can be obtained by a
perturbative expansion, justified in view of relative size of the SM
lepton masses and the much heavier ones 
of the vector-like leptons ($M_{L,E}$). In this study, we used
$\frac{(yv, \, hv_X,\, kv_X)}{M_{L,E}}\ll1$ as the (small) expansion
parameters, and followed the algorithm prescribed in~\cite{Grimus:2000vj}. 
Up to third order in the perturbation series, we thus obtain
\begin{equation}\label{eq:def:UellL}
  U_L = \begin{pmatrix}
          1 - \frac{\lambda_L^2 v_X^2}{4 M_L^2} & \frac{\lambda_L
            v_X}{\sqrt{2} M_L} - \frac{\lambda_L^3
            v_X^3}{4\sqrt{2}M_L^3} & \frac{(k \lambda_L M_E + h
            \lambda_L M_L + \lambda_E M_E y)v v_X}{2 M_E^3}\\ 
          \frac{\lambda_L^3 v_X^3}{4\sqrt{2}M_L^3} - \frac{\lambda_L
            v_X}{\sqrt{2}M_L} & 1 - \frac{\lambda_L^2 v_X^2}{4 M_L^2}
          & \frac{(k M_E M_L + h(M_E^2 + M_L^2))v}{\sqrt{2}M_E^3} \\ 
          \frac{(h \lambda_L M_E - \lambda_E M_L y)v v_X}{4 M_E^3} & -
          \frac{(k M_E M_L + h(M_E^2 + M_L^2))v}{\sqrt{2}M_E^3} & 1 
        \end{pmatrix}
\end{equation}
and
\begin{equation}\label{eq:def:UellR}
  U_R = \begin{pmatrix}
          1 - \frac{\lambda_E^2 v_X^2}{4 M_E^2} & \frac{\lambda_L v
            v_X}{2M_L^2} - \frac{\lambda_E(k M_E M_L + h(M_E^2 +
            M_L^2))v v_X}{2M_E^3 M_L} & \frac{\lambda_E
            v_X}{\sqrt{2}M_E} - \frac{\lambda_E^3 v_X^3}{4\sqrt{2}
            M_E^3}\\ 
          \frac{(h \lambda_E M_L - \lambda_L M_E y)v v_X}{2 M_E M_L^2}
          & 1 & \frac{(h M_E M_L + k(M_E^2 +
            M_L^2))v}{\sqrt{2}M_E^3}\\ 
          \frac{\lambda_E^3 v_X^3}{4\sqrt{2} M_E^3} - \frac{\lambda_E
            v_X}{\sqrt{2} M_E} & -\frac{(h M_E M_L + k(M_E^2 +
            M_L^2))v}{\sqrt{2}M_E^3} & 1 - \frac{\lambda_E^2 v_X^2}{4
            M_E^2} 
        \end{pmatrix}\,\text.
\end{equation}

\bigskip
Concerning the neutral leptons, the symmetric (Majorana) mass
matrix can be written as 
\begin{eqnarray}
   \mathcal L^{\nu}_{\text{mass}}&=& \begin{pmatrix}\nu^T&  N^{c\:T}&
     L^{0\,T}&  L^{0\:c\:T}\end{pmatrix}_L C^{-1} 
  \cdot M_{\nu}\cdot
  \begin{pmatrix}\nu\\N^c\\  L^{0} \\L^{0\:c}\end{pmatrix}_L\nonumber\\
  &=& \begin{pmatrix}\nu^T&  N^{c\:T}& L^{0\,T}&  L^{0\:c\:T}\end{pmatrix}_L C^{-1}
  \begin{pmatrix}
    0 & y_\nu \frac{v}{\sqrt{2}} & 0 &\lambda_L \frac{v_{X}}{\sqrt{2}}\\
    y_\nu \frac{v}{\sqrt{2}} & y_M \frac{v_{X}}{\sqrt{2}} & 0 & 0\\
    0 & 0 & 0 & M_L\\
    \lambda_L \frac{v_{X}}{\sqrt{2}} & 0 & M_L & 0\\
  \end{pmatrix}
  \begin{pmatrix}\nu\\N^c\\  L^{0} \\L^{0c}\end{pmatrix}_L\,\text,
  \label{eqn:numass}
\end{eqnarray}
in which each entry again corresponds to a $3\times3$ block
matrix. Following the same perturbative approach, and in this case up
to second order in perturbations of $\frac{y_\nu v}{y_M v_X},
\:\frac{y_\nu v}{M_L},\: \frac{y_M v_X}{M_L}\ll 1$, 
the mass matrix of Eq.~\eqref{eqn:numass}
can be block-diagonalised via a single unitary rotation
\begin{equation}
  M_\nu^\text{diag} = \tilde U_\nu^T\, M_\nu \,\tilde U_\nu\,,
\end{equation}
with 
\begin{equation}
  \tilde U_\nu = \begin{pmatrix}
           1 - \frac{\lambda_L^2 v_X^2}{4 M_L^2} - \frac{v^2
             y_\nu^2}{2 v_X^2 y_M^2} & \frac{v y_\nu}{v_X y_M} &
           \frac{\lambda_L v_X}{2 M_L} & \frac{\lambda_L v_X}{2
             M_L}\\ 
           -\frac{v y_\nu}{v_X y_M} & 1 -\frac{v^2 y_\nu^2}{2 v_X^2
             y_M^2} & 0 & 0\\ 
           -\frac{\lambda_L v_X}{\sqrt{2}M_L} & -\frac{\lambda_L v
             y_\nu}{\sqrt{2} M_L y_M} & \frac{1}{\sqrt{2}} -
           \frac{\lambda_L^2 v_X^2}{4\sqrt{2}M_L^2} &
           \frac{1}{\sqrt{2}} - \frac{\lambda_L^2
             v_X^2}{4\sqrt{2}M_L^2}\\ 
           0&0&-\frac{1}{\sqrt{2}} & \frac{1}{\sqrt{2}}
          \end{pmatrix}
          \,\text.
\end{equation}
We notice that the light (active) neutrino masses are generated via a
type-I seesaw mechanism 
\cite{Minkowski:1977sc,GellMann:1980vs,Yanagida:1979as,Mohapatra:1979ia,Schechter:1980gr,Mohapatra:1980yp,Schechter:1981cv},
relying on the Majorana mass term of the singlet right-handed 
neutrinos, $\sim v_X y_M/\sqrt 2$, which is dynamically generated upon
the breaking of $U(1)_{B-L}$; contributions from the vector-like
neutrinos arise only at higher orders and can therefore be safely neglected.
Up to second order in the relevant expansion parameters,
one then finds for the light (active) neutrino masses
\begin{equation}
	m_{\nu} \simeq -\frac{y_\nu^2 v^2}{v_X y_M}\,.
\end{equation}
As already mentioned, we work under the assumption that with the
exception of the neutrino Yukawa couplings $y_\nu$, all other couplings are
diagonal in generation space. Thus, the entire flavour structure
at the origin of leptonic mixing is encoded in the Dirac mass matrix
$(\propto v y_\nu)$, which can be itself diagonalised by a unitary matrix
$U_P$ as  
\begin{equation}
	\hat y_\nu \,= \,U_P^T\, y_\nu\, U_P\,.
\end{equation}
The full diagonalisation of the $12\times12$ neutral lepton mass matrix is
then given by 
\begin{equation}
	U_\nu \,= \,\tilde U_\nu \,\mathrm{diag}(U_P, \mathbb{1}, \mathbb{1}, \mathbb{1})\,,
\end{equation}
in which $\mathbb{1}$ denotes a $3\times3$ unity matrix.
In turn, this allows defining the leptonic charged current interactions as
\begin{equation}
  \mathcal L_{W^\pm} \,=\, -\frac{g}{\sqrt{2}} \,W_\mu^- \,
  \sum_{\alpha=e,\,\mu,\,\tau}\sum_{i = 1}^{9}\sum_{j = 1}^{12}
  \bar\ell_i \,(U_L^\dagger)_{i\,\alpha}\,\gamma^\mu \,P_L\,
  (U_\nu)_{\alpha\,j}\,\nu_j + \mathrm{H.c.}\,, 
\end{equation}
in which we have explicitly written the sums over flavour and mass
eigenstates (9 charged lepton mass eigenstates, and 12 neutral
states). 
The (not necessarily unitary~\cite{Xing:2007zj,Blennow:2016jkn,Fernandez-Martinez:2016lgt,Escrihuela:2015wra,Hati:2019ufv}) Pontecorvo–Maki–Nakagawa–Sakata (PMNS) matrix corresponds to
the upper $3\times 3$ block of $\sum_\alpha
(U_L^\dagger)_{i\,\alpha}(U_\nu)_{\alpha\,j}$ (i.e. $i,j=1,2,3$,
corresponding to the lightest, mostly SM-like states of both charged
and neutral lepton sectors).

\subsection{New neutral current interactions: $\pmb{Z^\prime}$ and
  $\pmb{h_X}$}\label{section:newneutralcurrent}

Having obtained all the relevant elements to characterise the lepton
and gauge sectors, we now address the impact of the additional fields
and modified couplings to the new neutral currents, in particular
those mediated by the light $Z^\prime$, which will be the key
ingredients to address the anomalies here considered. 
The new neutral currents mediated by the $Z^\prime$ boson, $ i
Z^\prime_\mu J_{Z^\prime}^\mu $
can be expressed as
\begin{align}
  J_{Z^\prime}^\mu \,=  \,e  \,\bar f_i  \,\gamma^\mu \,
  \left(\varepsilon^V_{ij} + \gamma^5  \,\varepsilon^A_{ij}\right) \,  f_j \, ,
	\label{eq:vector-current}
\end{align}
in which $f$ denotes a SM fermion (up- and down-type quarks, charged
leptons, and neutrinos) and the coefficients $\varepsilon^{V,A}_i$ are
the effective couplings in units of $e$. For the up- and down-type
quarks ($f=u,d$) the axial part of the $Z^\prime$ coupling formally
vanishes, $\varepsilon^{A}_q=0$, while the vector part is given by 
\begin{eqnarray}\label{eq:epsq}
  \varepsilon^{V}_{qq} &=&
  \varepsilon \,Q_q + \varepsilon_{B-L} \,Q_q^{B-L}\,\text.
\end{eqnarray}
On the other hand, and due to the mixings with the vector-like fermions,
the situation for the lepton sector is different: the
modified left- and right-handed couplings for the charged leptons
now lead to mixings between different species, as cast below
(for a given generation)
\begin{eqnarray}
  g_{Z^\prime, \,L}^{\ell_a \ell_b} &=&  \sum_{c=1,2,3}
  \left(\varepsilon \,Q_c + \varepsilon_{B-L} \,Q_c^{B-L}
  \right)(U_L^\dagger)^{ac}\,U_L^{cb}\, , 
  \label{eq:epslL}\\ 
  g_{Z^\prime,\, R}^{\ell_a \ell_b} &=&  \sum_{c=1,2,3}
  \left(\varepsilon \,Q_c + \varepsilon_{B-L} \,Q_c^{B-L}
  \right)(U_R^\dagger)^{ac}\,U_R^{cb}\,\text.
  \label{eq:epslR}
\end{eqnarray}
In the above, the indices $a,\,b,\,c$ refer to the 
mass eigenstates of different species:
the lightest one ($a,b,c=1$) corresponds to the (mostly)
SM charged lepton, while the two heavier ones (i.e. $a,b,c=2,3$)
correspond to the isodoublet and isosinglet heavy vector-like
leptons.
This leads to the following vectorial and axial couplings
\begin{eqnarray}\label{eq:epsl}
  g^{V}_{\ell_a \ell_b} = \frac{1}{2} \left({g_{Z^\prime, \,L}^{\ell_a
      \ell_b}+ g_{Z^\prime,\, R}^{\ell_a \ell_b}}\right)\, , \quad
  g^{A}_{\ell_a \ell_b} = \frac{1}{2} \left({g_{Z^\prime, \,R}^{\ell_a
      \ell_b}- g_{Z^\prime,\, L}^{\ell_a \ell_b}}\right)\,\text. 
\end{eqnarray} 
Similarly, the new couplings to the (Majorana) neutrinos are given by
\begin{eqnarray}\label{eq:epsnu}
   g^{V}_{\nu_a \nu_b}  &=&\sum_{c} \varepsilon_{B-L}\,
   \mathrm{Im}\left(Q_c^{B-L} \,(U_\nu^\ast)^{ca}\, U_\nu^{c
     b}\right)\,\text,\\ 
   g^{A}_{\nu_a \nu_b}  &=&-\sum_{c} \varepsilon_{B-L}\,
   \mathrm{Re}\left(Q_c^{B-L} \,(U_\nu^\ast)^{ca} \,U_\nu^{c
     b}\right)\,\text.\label{eqn:gAnu}
\end{eqnarray}
Note that the vector part of the couplings vanishes for
$\nu_a = \nu_b$ (with $a,b=1,2$),
which corresponds to the Majorana 
mass eigenstates with purely Majorana masses (cf. Eq.~\ref{eqn:numass}).  
For the lightest (mostly SM-like) physical states ($a,b=1$) one has
\begin{eqnarray}
	\varepsilon^A_{\nu_\alpha\nu_\alpha} =
        - g_{Z^\prime,\,L}^{\nu_\alpha\nu_\alpha} &\simeq& \,
        \varepsilon_{B-L} \left(1 - \frac{\lambda_{L\:\alpha}^2
          v_{X}^2}{M_{L\:\alpha}^2} \right)\label{eqn:nunu}\,\text,\\ 
  g_{Z^\prime,\,L}^{\ell_\alpha\ell_\alpha} & \simeq&  - \varepsilon
  + \left( \frac{\lambda_{L\:\alpha}^2 v_{X}^2}{M_{L\:\alpha}^2} -
  1\right)\,\varepsilon_{B-L}\label{eqn:eeL}\,\text,\\ 
  g_{Z^\prime,\,R}^{\ell_\alpha\ell_\alpha} & \simeq& - \varepsilon +
  \left(\frac{\lambda_{E\:\alpha}^2 v_{X}^2}{M_{E\:\alpha}^2} -
  1\right)\varepsilon_{B-L}\,\text,\label{eqn:eeR}\\ 
	\varepsilon^{V}_{\ell_\alpha\ell_\alpha} &\simeq&  -
        \varepsilon +\frac{1}{2}\left(\frac{\lambda_{L\,\alpha}^2
          v_X^2}{M_{L\,\alpha}^2} + \frac{\lambda_{E\,\alpha}^2
          v_X^2}{M_{E\,\alpha}^2} - 2 \right)\varepsilon_{B-L}\,, \\ 
	\varepsilon^{A}_{\ell_\alpha\ell_\alpha} &\simeq&
        \frac{1}{2}\left(\frac{\lambda_{E\,\alpha}^2
          v_X^2}{M_{E\,\alpha}^2} - \frac{\lambda_{L\,\alpha}^2
          v_X^2}{M_{L\,\alpha}^2}  \right)\varepsilon_{B-L}\,, 
\end{eqnarray}
in which the subscript $\alpha\in\{e, \,\mu,\,\tau\}$ now
explicitly denotes the SM lepton flavour.
Note that flavour changing (tree-level) couplings are absent by
construction, as a consequence of having imposed strictly diagonal
couplings and masses ($\lambda_{L,\,E}$, $M_{L,\,E}$) for the
vector-like states. The ``cross-species couplings'' are defined in
Eqs.~\eqref{eq:epslL}, \eqref{eq:epslR} and \eqref{eqn:gAnu}. 

\bigskip
Finally, the scalar and pseudoscalar couplings of $h_X$ to the
charged leptons (SM- and vector-like species) can be conveniently
expressed as 
\begin{equation}
	\frac{v_X}{\sqrt{2}}\,g_S \,=\, m^\ell_\text{diag} -
        \frac{1}{2}\left(C_{LR} + C_{RL}\right) 
\end{equation}
and
\begin{equation}
	\frac{v_X}{\sqrt{2}}\,g_P \,=\,  \frac{1}{2}\left(C_{LR} -
        C_{RL}\right)\,\text, 
\end{equation}
where
\begin{equation}
	C_{LR} = (C_{RL})^\dagger =
	U_L^\dagger
  	\begin{pmatrix}
  	\frac{y v}{\sqrt{2}} & 0 & 0\\
  	0&M_L&\frac{h v}{\sqrt{2}}\\
  	0&\frac{k v}{\sqrt{2}} & M_E
  	\end{pmatrix}
 	 U_R\,\text,
\end{equation}
with $U_{L,R}$ as defined in Eqs.~(\ref{eq:def:UellL}, \ref{eq:def:UellR}).
Further notice that corrections to the tree-level couplings of the SM
Higgs and $Z$-boson appear only at higher orders in the
perturbation series of the mixing matrices, and are expected to be of
little effect.

\section{New physics
  contributions to the anomalous magnetic moments}\label{sec:g-2} 

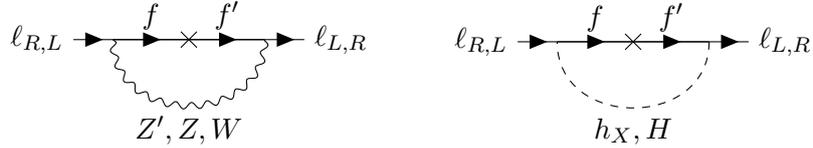
\begin{figure}[h!]
\begin{center}
\feynmandiagram [layered layout, horizontal=b to c] {
i1 [particle=\(\ell_{R,L}\)]
a -- [fermion] b
-- [fermion, edge label=\(f\)] c1
-- [fermion, edge label=\(f^\prime\)] c
-- [fermion] f1 [particle=\(\ell_{L,R}\)] d,
b -- [insertion={[style=black]0.5}] c,
c-- [photon, half left, edge label=\({Z^\prime, Z, W}\)] b,
}; \quad \quad 
\feynmandiagram [layered layout, horizontal=b to c] {
i1 [particle=\(\ell_{R,L}\)]
a -- [fermion] b
-- [fermion, edge label=\(f\)] c1
-- [fermion, edge label=\(f^\prime\)] c
-- [fermion] f1 [particle=\(\ell_{L,R}\)] d,
b -- [insertion={[style=black]0.5}] c,
c-- [scalar, half left, edge label=\({h_X, H}\)] b,
};
\caption{Illustrative Feynman diagrams for the one-loop contributions to
  $(g-2)_{e,\mu}$ induced by the new states and couplings (with a
  possible mass insertion inside the 
  loop or at an external leg). The internal states ($f,\,f^\prime$)
  are leptons; the photon can be
  attached to any of the charged fields.}
\label{fig:feyn}
\end{center}
\end{figure}

The field content of the model gives rise to new contributions to the
anomalous magnetic moments of the light charged leptons, in the
form of several one-loop diagrams mediated by the extra $Z^\prime$ and $h_X$
bosons, as well as the new heavy vector-like fermions, which can also
propagate in the loop. The new contributions are schematically
illustrated in Fig.~\ref{fig:feyn}. Notice that due to the potentially
large couplings, the contributions induced by the $Z^\prime$ or even $h_X$
can be dominant when compared to the SM ones.
%

Generic one-loop contributions generated by the exchange of a 
neutral vector boson (NV) and a negatively charged internal fermion,
$\Delta a_\ell^\text{NV}$,
can be expressed as~\cite{Jegerlehner:2009ry}
\begin{equation}
  \Delta a_\ell^\text{NV} \,=  \,
  \sum_{i}\left[\frac{g_V^{\ell\,i} \,{g_V^{\ell\,i}}^\ast}{4\pi^2} \,
    \frac{m_\ell^2}{m_B^2} \,F(\lambda, \rho_i) +
    \frac{g_A^{\ell\,i} \,{g_A^{\ell\,i}}^\ast}{4\pi^2} \,
    \frac{m_\ell^2}{m_B^2} \,F(\lambda, -\rho_i)\right]\,\text,
\end{equation}
with $\Delta a_\ell$ as defined in Eqs.~(\ref{Delta_amu},\ref{Delta_ae});  
$g_{V(A)}$ is the vector (axial-vector) coupling\footnote{The sum in
  Eq.~\eqref{mdmvcl} runs over 
  all fermions which have non-vanishing couplings to the external
  leptons, so that in general $i = 1,\,2,\,\dots 9$; however note that
  only fermions belonging to the same generation (but possibly of
  different species e.g., SM leptons and isosinglet or isodoublet
  vector-like leptons) have a non-vanishing entry.}  and $m_B$ is the mass
of the exchanged vector boson. The function $F(\lambda, \rho)$ is
defined as follows
\begin{equation}{\label{mdmvcl}}
  F(\lambda, \rho)  \,= \,
  \frac{1}{2}\int_0^1 \frac{2x \,(1-x) \,\left[x-2(1-\rho)\right] + \lambda^2
    \,x^2 \,(1-\rho)^2 \,(1+\rho -x)}{(1-x)(1-\lambda^2 x) +
    (\rho\lambda)^2  \,x}\, dx \,\text,
\end{equation}
in which $\rho_i = M_{f_i} / m_\ell$ with $M_{f_i}$ denoting the
internal fermion mass and with $\lambda = m_\ell / M_B$.

\noindent
Generic new contributions due to a neutral scalar mediator (NS),
$\Delta a_\ell^\text{NS}$, are given by
\begin{equation}
  \Delta a_\ell^\text{NS}\, =\,
  \sum_{i}\left(\frac{g_S^{\ell\,i}\,{g_S^{\ell\,i}}^\ast}{4\pi^2}\,
  \frac{m_\ell^2}{m_S^2}\,G(\lambda, \rho_i) +
  \frac{g_P^{\ell\,i}\,{g_P^{\ell\,i}}^\ast}{4\pi^2}\,
  \frac{m_\ell^2}{m_S^2}\,G(\lambda, -\rho_i)\right)\,\text,
\end{equation}
with
\begin{equation}
  G(\lambda,\rho) \,=\,
  \frac{1}{2}\int_{0}^{1} dx \frac{x^2\,(1 + \rho - x)}{(1-x)\,
    (1-\lambda^2 \,x) + (\rho\,\lambda)^2 \,x}\,\text,
\end{equation}
in which $g_{S(P)}$ denotes the scalar (pseudoscalar) coupling
and $m_S$ is the mass of the neutral scalar $S$. Note that the loop
functions of a vector or a scalar mediator have an overall positive
sign, whereas the contributions of axial and pseudoscalar mediators
are negative. This allows for a partial cancellation between
vector and axial-vector contributions, as well as
between scalar and pseudoscalar
ones, which are crucial
to explain the relative (opposite) signs of
$\Delta a_e$ and $\Delta a_\mu$. As expected, such cancellations
naturally rely on the interplay of the $Z^\prime$
and $h_X$ couplings. 


\section{Explaining the anomalous IPC in $\pmb{^8}$Be}\label{sec:IPCcon} 
We proceed to discuss how the presence of a light $Z^\prime$ boson and
the modified neutral currents can successfully address the  
internal pair creation anomaly in $^8$Be atoms\footnote{As already
  mentioned in the Introduction, in~\cite{Krasznahorkay:2019lyl} it
  has been reported that a peak in the electron-positron pair angular
  correlation was observed in the electromagnetically forbidden
  $M0$ transition depopulating the $21.01$ MeV $0^{-}$ state in
  $^4$He, which could be explained by the creation and subsequent decay of a
  light particle in analogy to $^8$Be. However, in the absence of any fit
  for normalised branching fractions we will not include this in our
  analysis.}.

Firstly, let us consider one of the quantities which is extremely
relevant for the IPC excess - the width of the $Z^\prime$ decay into
a pair of electrons.   
At tree level, the latter is given by 
\begin{align}\label{eqn:Vff}
\Gamma(Z^\prime\to e^{+} e^{-}) \,= \,\left(| \varepsilon_{ee}^V|^2 + |
\varepsilon_{ee}^A|^2 \right) \frac{\lambda^{1/2}(m_{Z^\prime}, m_e ,
  m_{e})}{24 \,\pi \,m_{Z^\prime}} \,,
\end{align}
where the K\"all\'en function is defined as
$\lambda(a,b,c) = \left(a^{2} - \left(b-c\right)^{2} \right)\left(a^{2} -
\left(b+c\right)^{2} \right)$.

In what follows we discuss the bounds on the $Z^\prime$ which are directly
connected with an explanation of the $^8$Be anomaly. 
A first bound on the couplings of the $Z^\prime$ can be obtained from
the requirement that the $Z^\prime$ be sufficiently short lived for its
decay to occur inside the Atomki spectrometer, whose length is 
$\cal{O}$(cm). This gives rise to 
a lower bound on the couplings of the $Z^\prime$ to electrons 
\begin{align}\label{eq:4.2}
|\varepsilon^V_{ee}|
\gtrsim
{1.3 \times 10^{-5}} \sqrt{\text{BR}(Z^\prime\to e^+e^-)} \, .
\end{align}

The most important bounds clearly arise from the requirement that
$Z^\prime$ production (and decay) complies with the (anomalous)
data on the electron-positron angular correlations for the $^8$Be
transitions. 
We begin by recalling that the relevant quark (nucleon) couplings
necessary to explain the anomalous IPC in $^8$Be can be determined
from a combination of
the best fit value for the normalised branching fractions
experimentally measured. This is done here for both the cases of isospin conservation and breaking.

\begin{figure}[hbt!]
\begin{center}
\mbox{    \includegraphics[width = 0.48 \textwidth]{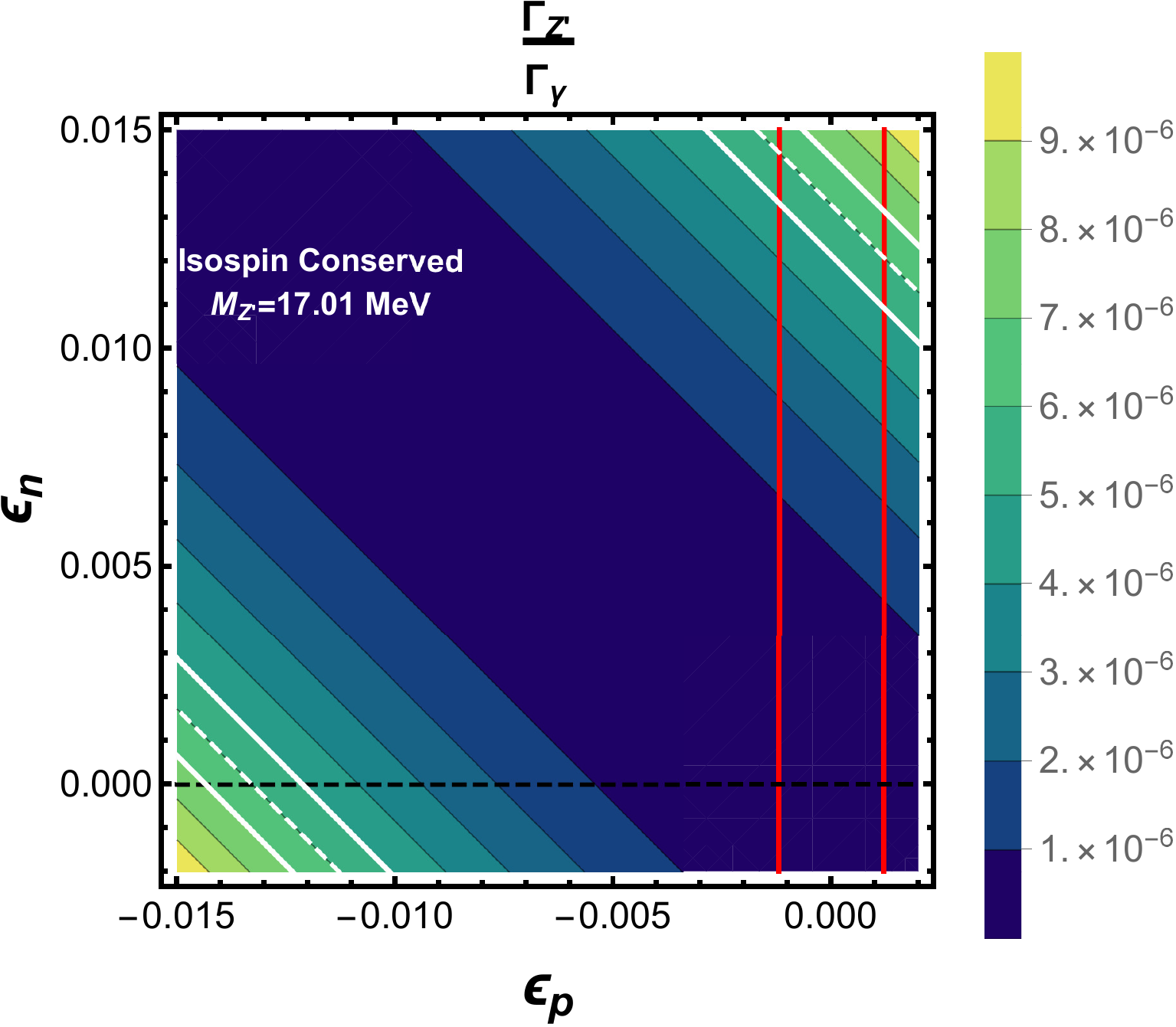} \quad\quad 
     \includegraphics[width = 0.48 \textwidth]{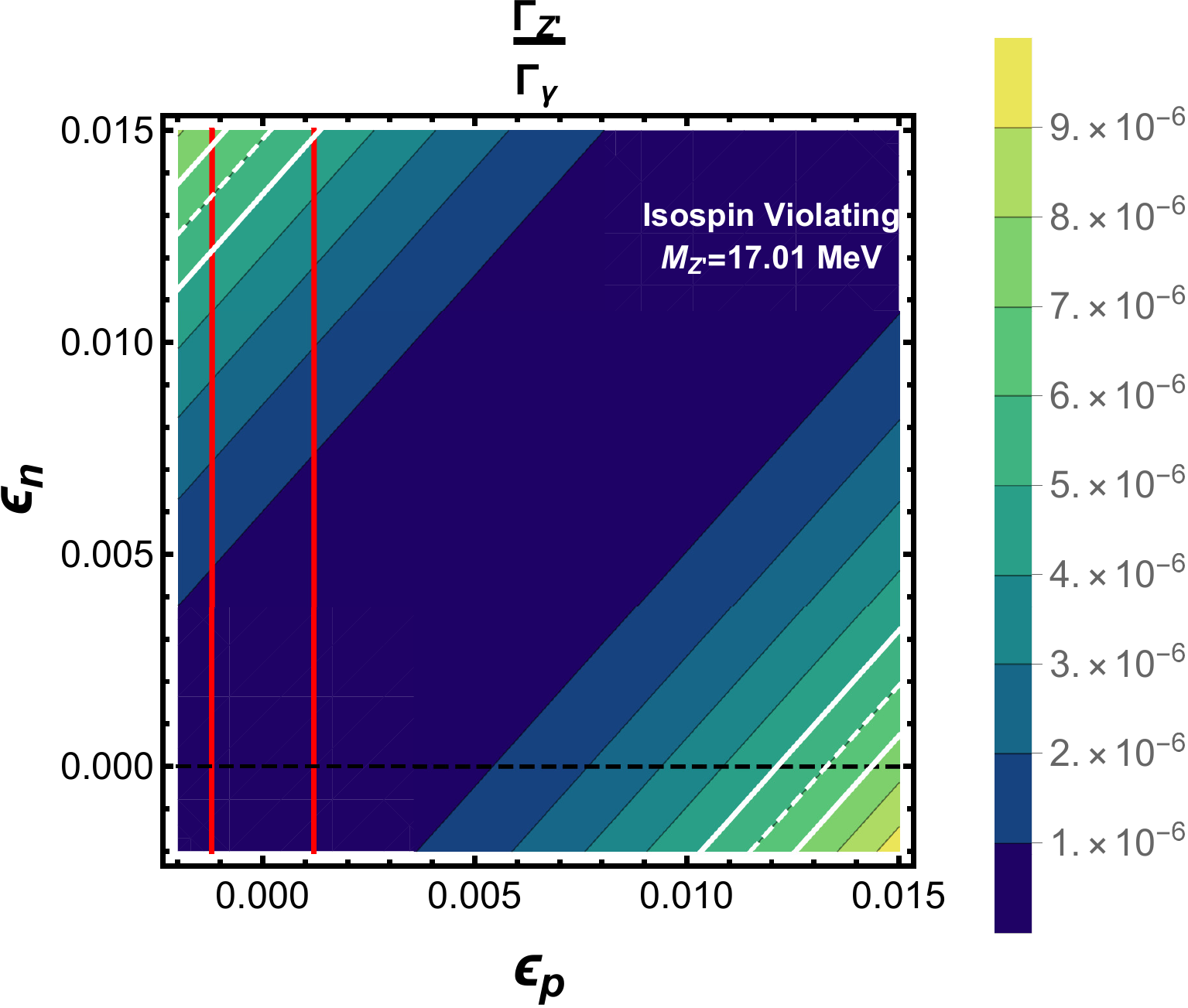}}
     \vspace*{3mm}\\
\mbox{		  \includegraphics[width = 0.48 \textwidth]{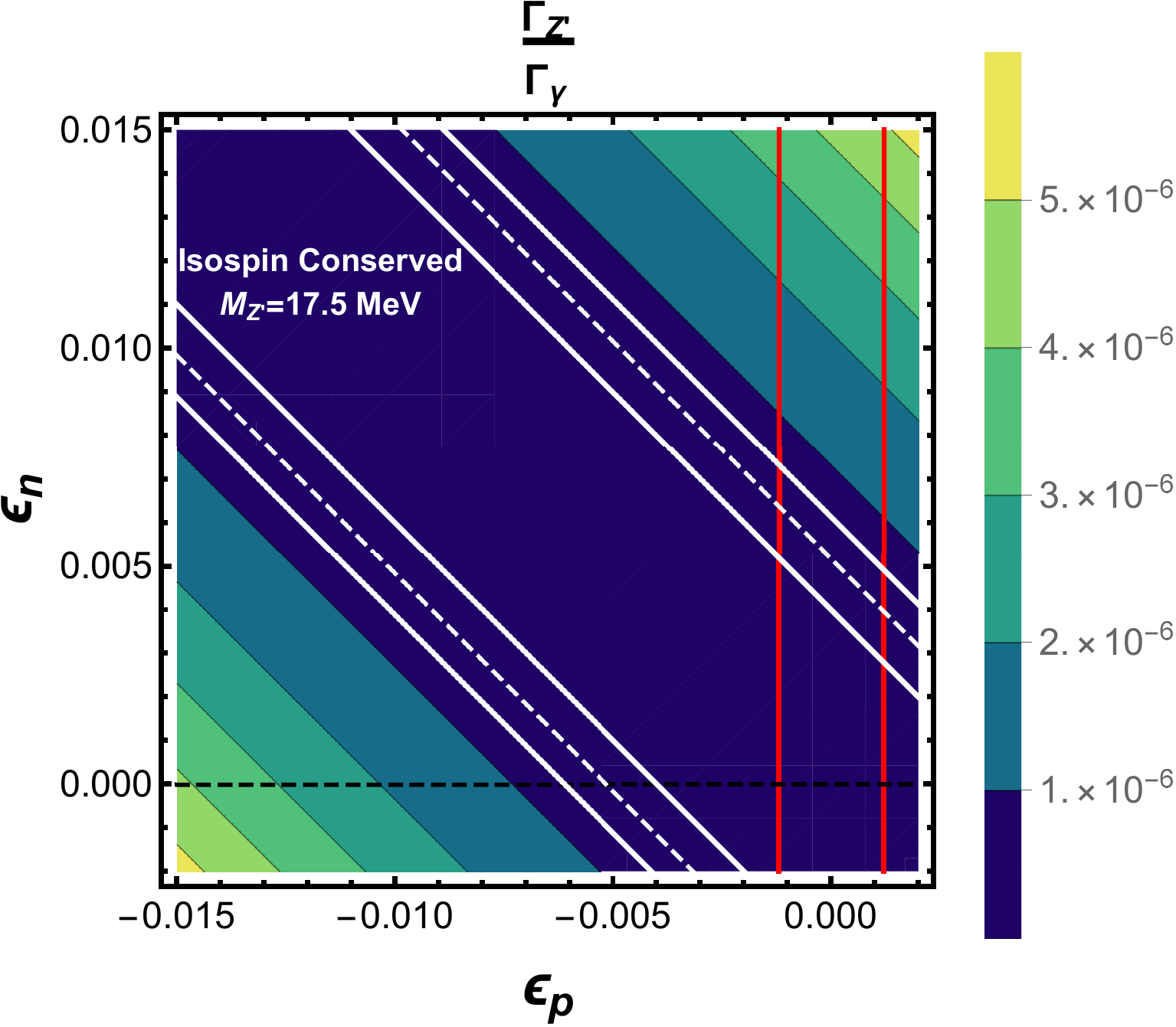} \quad \quad 
       \includegraphics[width = 0.48 \textwidth]{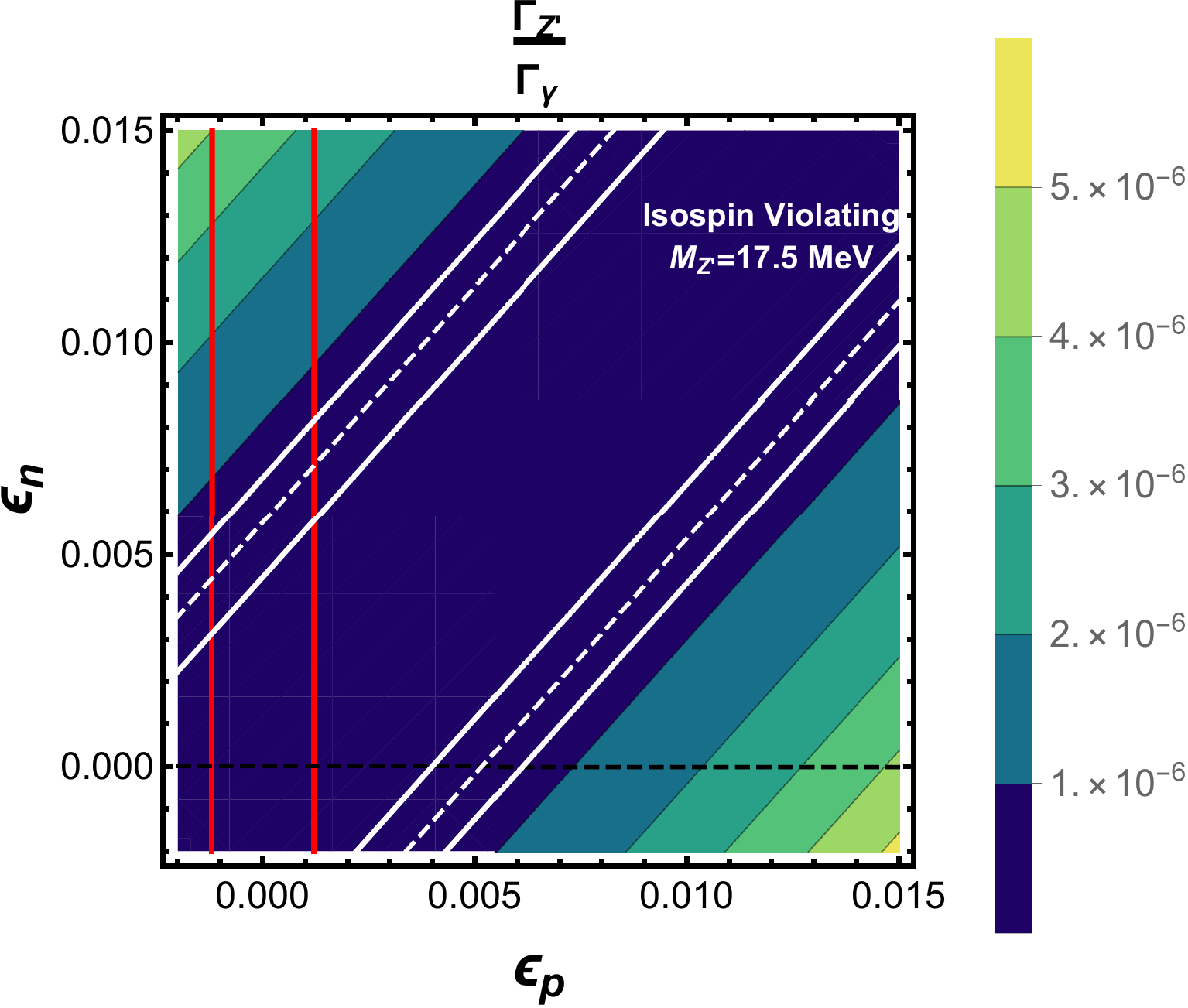}}
       \caption{On the left (right) panels,
         contour plots of the ratio $\Gamma_{Z^\prime}/\Gamma_{\gamma}$
         (see Eq.~(\ref{eq:def:rationGammaZpgamma})
         for the isospin conserving (violating) limit. The white dashed
and solid lines correspond to the best fit and to the 1$\sigma$ interval for
the experimental best fit values for $\Gamma_{Z^\prime}/\Gamma_{\gamma}$,
under the assumption BR$(Z^\prime\to e^{+} e^{-})=1$.
The region between the two red
vertical lines corresponds to the viable protophobic region of the
parameter space, as allowed by NA48/2 constraints, while the horizontal
dashed line corresponds to the pure dark photon limit.
On both upper panels we have taken $m_{Z^\prime}$=17.01~MeV, as well as
the experimental best fit value 
$\Gamma_{Z^\prime}/\Gamma_{\gamma}=6 (1) \times 10^{-6}$ (following the fit
values reported in~\cite{Krasznahorkay:2018snd}). The lower
panels illustrate the case in which $m_{Z^\prime}$=17.5~MeV, for an
experimental best fit value $\Gamma_{Z^\prime}/\Gamma_{\gamma}=0.5 (0.2)
\times 10^{-6}$, in agreement with the values quoted in~\cite{Feng:2016ysn}
(for which we have taken a conservative estimate of the error in 
$\Gamma_{Z^\prime}/\Gamma_{\gamma} \sim 0.2 \times 10^{-6}$, following the
uncertainties of~\cite{Krasznahorkay:2018snd}). 
}
\label{fig:lim}
\end{center}
\end{figure}

\paragraph{Isospin conservation}
In the isospin conserving
limit, the normalised branching fraction
\begin{equation}\label{eq:def:rationGammaZpgamma}
\frac{\Gamma({^8\text{Be}^*} \to
  {^8\text{Be}}\, Z^\prime)}{
  \Gamma({^8\text{Be}^*} \to {^8\text{Be}}\, \gamma)}
\equiv
\frac{\Gamma_{Z^\prime}}{\Gamma_{\gamma}}
\end{equation}  
is a particularly convenient observable because the relevant nuclear
matrix elements cancel in the ratio, giving 
\begin{align}\label{eq:4.3}
	\frac{
	\Gamma({^8\text{Be}^*} \rightarrow {^8\text{Be}}+Z^\prime)
	}{
	\Gamma({^8\text{Be}^*} \rightarrow {^8\text{Be}}+\gamma)
	}
	&=
	(\varepsilon^V_p+\varepsilon^V_n)^2
	\frac{|\mathbf{k}_{Z^\prime}|^3}{|\mathbf{k}_\gamma|^3}
	=
	(\varepsilon^V_p+\varepsilon^V_n)^2
	\left[1 - \left(\frac{m_{Z^\prime}}{18.15\text{ MeV}}
          \right)^2\right]^{3/2}
	\, ,
\end{align}
in which $\varepsilon^V_p = 2\,\varepsilon^V_{uu} + \varepsilon^V_{dd}$
and $\varepsilon^V_n = \varepsilon^V_{uu} + 2 \,\varepsilon^V_{dd}$.
The purely vector quark currents (see Eq.~\eqref{eq:vector-current}) are
expressed as 
\begin{align}
	J_{Z^\prime}^{\mu \, ({\rm q})}
	= \sum_{i=u,d} \varepsilon^V_{ii} e J_i^\mu\,  \quad 
	(J_i^\mu = \bar q_i \gamma^\mu q_i) \, .
	\label{eq:quark-vector-current}
\end{align}
The cancellation of the nuclear matrix elements in the ratio of
Eq.~\eqref{eq:4.3} can be understood as described below. Following the
prescription of Ref.~\cite{Feng:2016ysn}, it is convenient to parametrise
the matrix element for nucleons in terms of the Dirac and Pauli form
factors $F^{Z^\prime}_{1,p} (q^2)$ and $F^{Z^\prime}_{2,p}
(q^2)$~\cite{Perdrisat:2006hj}, so that the proton matrix element can be
written as 
\begin{align}
J_p^\mu \equiv
\langle p(k') | J_{Z^\prime}^{\mu \, ({\rm q})} | p(k) \rangle  & =
e \,\overline{u}_p(k')  \left\{ F^{Z^\prime}_{1,p} (q^2)
\,\gamma^\mu + F^{Z^\prime}_{2,p} (q^2)
\,\sigma^{\mu \nu} \,\frac{q_\nu} {2 M_p }\right\} u_p (k) \,.
\end{align}
Here  $| p(k) \rangle$ corresponds to a proton state and $u_p (k)$
denotes the spinor corresponding to a free proton. The nuclear
magnetic form factor is then given by $G_{M,p}^{Z^\prime}(q^2) =
F^{Z^\prime}_{1,p} (q^2) + F^{Z^\prime}_{2,p}
(q^2)$~\cite{Perdrisat:2006hj,Yennie:1957,Ernst:1960zza,Hand:1963zz}. The
nucleon currents can be combined to obtain the isospin currents as 
\begin{equation}
	J_{0}^\mu \,=\, J_p^\mu \,+\, J_n^\mu\,, \quad \quad
	J_{1}^\mu \,=\, J_p^\mu \,-\, J_n^\mu
	\,.
	\label{eq:isospin-current}
\end{equation}
In the isospin conserving limit,
$\langle {^8\text{Be}} | J_1^\mu |{^8\text{Be}^*}\rangle = 0$, since
both the exited and the ground state of $^8\text{Be}$ are
isospin singlets. Defining the $Z^\prime$ hadronic current as
\begin{align}
	J_{Z^\prime}^{\mu\, h} \,=\,
 \sum_{i=u,d} \varepsilon^V_{ii} \,e \,J_i^\mu
	= (2\,\varepsilon^V_{uu}+\varepsilon^V_{dd})\,e\, J_p^\mu
	+ (\varepsilon^V_{uu}+2\,\varepsilon^V_{dd})\,e\, J_n^\mu  \ ,
\end{align}
with $p, n$ denoting protons and neutrons, one obtains
\begin{align}
	\langle {^8\text{Be}} | J_{Z^\prime}^{\mu\, h} |{^8\text{Be}^*}\rangle
	&= \frac{e}{2} \,(\varepsilon_p + \varepsilon_n)
	\langle {^8\text{Be}} | J_0^\mu |{^8\text{Be}^*}\rangle\,,
	\label{eq:isospin:matrix:elements1}
	\\
	\langle {^8\text{Be}} | J_\text{EM}^\mu |{^8\text{Be}^*}\rangle
	&= \frac{e}{2}\,
	\langle {^8\text{Be}} | J_0^\mu |{^8\text{Be}^*}\rangle \, ,
	\label{eq:isospin-matrix-elements}
\end{align}
in which $\varepsilon_p = 2\,\varepsilon^V_{uu} + \varepsilon^V_{dd}$
and $\varepsilon_n = \varepsilon^V_{uu} + 2 \,\varepsilon^V_{dd}$. From
Eq.~\eqref{eq:isospin-matrix-elements} it follows that the relevant
nuclear matrix elements cancel in the normalised branching fraction
of Eq.~\eqref{eq:4.3} (in the isospin conserving
limit). Therefore,
using the best fit values for
the mass  $m_{Z^\prime}$=17.01~(16)~MeV~\cite{Krasznahorkay:2019lyl}, and 
the normalised branching
fraction $\Gamma_{Z^\prime}/\Gamma_{\gamma}=6 (1) \times 10^{-6}$,  
Eq.~\eqref{eq:4.3} leads to the following constraint
\begin{align}\label{eq:4.4.1}
  |\varepsilon^V_p + \varepsilon^V_n| \approx
	\frac{1.2 \times 10^{-2}}{\sqrt{\text{BR}(Z^\prime \to e^+e^-)}}
		\,.
\end{align}
On the top left panel of Fig.~\ref{fig:lim} we display the
 plane spanned by $\varepsilon_p$ vs. $\varepsilon_n$, for a hypothetical
$Z^\prime$ mass of $m_{Z^\prime}$=17.01~MeV, and for
the experimental best fit value
$\Gamma_{Z^\prime}/\Gamma_{\gamma}=6 (1) \times 10^{-6}$ (following the
most recent best fit values reported in~\cite{Krasznahorkay:2018snd}). 
Notice that a large departure of $|\varepsilon_p|$ from the
protophobic limit is excluded by NA48/2 constraints~\cite{Raggi:2015noa}, which are
depicted by the two red vertical lines. The region between the latter 
corresponds to the viable protophobic regime still currently
allowed. The horizontal dashed line denotes the limiting case of
a pure dark photon. 

\paragraph{Isospin breaking}
In the above discussion it has been implicitly assumed that the $^8$Be
states have a well-defined isospin; however, as extensively noted in the
literature~\cite{Barker:1966zza,Wiringa:2000gb,Pieper:2004qw,Wiringa:2013fia,Pastore:2014oda},
the $^8$Be states are in fact isospin-mixed. In order to take the latter
effects into account, isospin breaking in the electromagnetic
transition operators arising from the neutron--proton mass difference
was studied in detail in Ref.~\cite{Feng:2016ysn}, and found to have
potentially serious implications for the quark-level couplings
required to explain the $^8$Be signal. In what follows we summarise
the most relevant points, which will be included in the present study

\noindent
For a doublet of spin $J$, the
physical states (with superscripts $s_1$ and $s_2$) can be defined
as~\cite{Pastore:2014oda} 
\begin{equation}
  \Psi_J^{s_1} \,= \,\alpha_J \,\Psi_{J, T=0} \,+\, \beta_J\, \Psi_{J, T=1}\,,
  \quad\quad
  \Psi_J^{s_2} \,=\, \beta_J \,\Psi_{J, T=0} \,- \,\alpha_J\, \Psi_{J, T=1} \,,
  \label{mixing}
\end{equation}
in which the relevant mixing parameters $\alpha_J$ and $\beta_J$ can
be obtained by computing the widths of the isospin-pure states using
the  	
Quantum Monte Carlo (QMC) approach~\cite{Pastore:2014oda}. As pointed out
in~\cite{Feng:2016ysn}, this procedure may be used for the
electromagnetic transitions of isospin-mixed states as well. However,
the discrepancies with respect to the experimental results are
substantial,
even after including the meson-exchange currents in the relevant matrix
element~\cite{Pastore:2014oda}. To address this deficiency,
an isospin breaking effect was introduced
in the hadronic form factor of the electromagnetic transition operators
themselves in Ref.~\cite{Feng:2016ysn}
(following~\cite{Gardner:1995uq,Gardner:1995ya}). This has led to 
changes in the relative strength of the isoscalar and isovector
transition operators which appear as a result of isospin-breaking in
the masses of isospin multiplet states, e.g. the nonzero
neutron-proton mass difference. The isospin-breaking contributions
have been 
incorporated through the introduction of a spurion, which regulates
the isospin-breaking effects within an isospin-invariant framework
through a ``leakage'' parameter (controlled by non-perturbative
effects). The ``leakage'' parameter
is subsequently determined by matching the resulting
$M1$
transition rate of the 17.64~MeV decay of $^8$Be with its experimental
value, using the matrix elements of Ref.~\cite{Pastore:2014oda}. This
prescription leads to the corrected ratio of partial
widths~\cite{Feng:2016ysn},  
\begin{equation}
\frac{\Gamma(^8\text{Be}^* \rightarrow {^8\text{Be}}+Z^\prime)}{
\Gamma(^8\text{Be}^* \rightarrow {^8\text{Be}}+\gamma)}
=| 0.05 \, (\varepsilon^V_p + \varepsilon^V_n)   + 0.95 \,
(\varepsilon^V_p - \varepsilon^V_n) |^2 
\left[1 - \left(\frac{m_{Z^\prime}}{18.15\text{ MeV}}\right)^2\right]^{3/2} \, ,
\label{eq:IVH}
\end{equation}
and consequently, to new bounds on the relevant quark (nucleon)
couplings necessary to explain the anomalous IPC in $^8$Be.  
On the upper right panel of Fig.~\ref{fig:lim}, we display the
isospin-violating scenario of Eq.~\eqref{eq:IVH}, in the
$\varepsilon_p$ vs. $\varepsilon_n$ plane, for $m_{Z^\prime}$=17.01~MeV and
for the experimental best fit value 
$\Gamma_{Z^\prime}/\Gamma_{\gamma}=6 (1) \times
10^{-6}$~\cite{Krasznahorkay:2018snd}.
A comparison with the case of isospin conservation (upper left plot)
reveals a~$15\%$
modification with respect to 
the allowed protophobic range of $\varepsilon_n$ in
the isospin violating case.

\bigskip
Other than the best fit values for the mass of the mediator
and normalised branching fraction for the
predominantly isosinglet $^8$Be excited state with an excitation energy
18.15~MeV (here denoted as $^8\text{Be}^*$), it is 
important to take into account the IPC null results
for the predominantly isotriplet excited
state ($^8\text{Be}^{*'}$), as emphasised in~\cite{Gulyas:2015mia}. 
In particular, in the presence of a finite
isospin mixing, the latter IPC null result would call for 
a kinematic suppression, thus implying a larger preferred mass
for the $Z^\prime$, in turn leading to a large phase space suppression.
This may translate into (further) significant changes for
the preferred quark (nucleon) couplings
to the $Z^\prime$ (corresponding to a heavier $Z^\prime$, 
and to significantly smaller
normalised branching fractions when compared to the preferred fit
reported in~\cite{Krasznahorkay:2019lyl}).
Considering the benchmark value\footnote{Since no public results are
  available to the best of our knowledge, we use the values quoted
  from a private communication in~\cite{Feng:2016ysn}.} 
$\Gamma_{Z^\prime}/\Gamma_{\gamma}=0.5 \times 10^{-6}$~\cite{Feng:2016ysn}, 
we obtain the following constraint in the isospin conserving limit,
\begin{align}\label{eq:4.4}
  |\varepsilon^V_p + \varepsilon^V_n| \approx
 \frac{(3-6) \times 10^{-3}}{\sqrt{\text{BR}(Z^\prime\to e^+e^-)}}
   \, . 
\end{align}
Leading to the above limits, we have used a conservative estimate for
the error in $\Gamma_{Z^\prime}/\Gamma_{\gamma}$ ($\sim 0.2 \times 10^{-6}$)
following the quoted uncertainties in~\cite{Krasznahorkay:2018snd}.
In Fig.~\ref{fig:lim}, the bottom panels illustrate
the relevant parameter space for the isospin conserving and isospin
violating limits (respectively left and right plots).

To summarise, it is clearly important to further improve
the estimation of nuclear isospin violation, and perform
more accurate fits for the null result of IPC in $^8 \text{Be}^{*'}$
(in addition to the currently available fits for the
predominantly isosinglet $^8$Be excited state). This will allow
determining the ranges for the bounds on the relevant quark (nucleon)
couplings of the $Z^\prime$ necessary to explain the anomalous IPC in
$^8$Be.
However, in view of the guesstimates discussed here, in our
numerical analysis we will adopt conservative ranges for
different couplings (always under the assumption $\text{BR}(Z^\prime \to e^+
e^-)= 1$),
\begin{eqnarray}
| \varepsilon^V_n | &\;=\;& (2-15) \times 10^{-3}\, ,
\label{eqn:epsn}
\\
| \varepsilon^V_p | &\;\lesssim \; & 1.2 \times 10^{-3}\, .
\label{eqn:epsp}
\end{eqnarray}
%


\section{Phenomenological constraints on neutral (vector and axial) couplings}
\label{sec:phenocon}

If, and as discussed in the previous section,
the new couplings of fermions to the light $Z^\prime$ must satisfy
several requirements to explain the anomalous IPC in $^8$Be, there are
extensive constraints arising from various experiments, both regarding
leptonic and hadronic couplings. In this section, we collect the
most important ones, casting them in a model-independent way,  
and subsequently summarising the results of
the new fit carried for the case of light Majorana 
neutrinos (which is the case in the model under consideration).

\subsection{Experimental constraints on a light $\pmb{Z^\prime}$
  boson}\label{sec:phenocon:Zprime} 
The most relevant constraints arise
from negative $Z^\prime$ searches in beam dump
experiments, dark photon bremsstrahlung and production,
parity violation, and neutrino-electron scattering.

\paragraph{Searches for $\pmb{Z^\prime}$ in electron beam dump experiments}

\noindent
The non-observation of a $Z^\prime$ in experiments such
as SLAC E141, Orsay and NA64~\cite{Banerjee:2018vgk}, as well as 
searches for dark photon bremsstrahlung from electron and nuclei
scattering, can be interpreted in a two-fold way:
(i) absence of $Z^\prime$ production due to excessively feeble couplings;
(ii) excessively rapid $Z^\prime$ decay, occurring even prior to the dump.
Under assumption (i) (i.e. negligible production), one finds the following bounds
\begin{equation}
	{\varepsilon^V_{ee}}^2 + {\varepsilon^A_{ee}}^2 < 1.1\times10^{-16}\,,
\end{equation}
while (ii) (corresponding to fast decay) leads to 
\begin{equation}
	\sqrt{{|\varepsilon^V_{ee}|}^2 + {|\varepsilon^A_{ee}|}^2}
	 \gtrsim 4\times10^{-4}\: \sqrt{\text{BR}(Z^\prime\to e^+e^-)}\,.
	 \label{eqn:dump}
\end{equation}

\paragraph{Searches for dark photon production}

\noindent
A bound can also be obtained from the KLOE-2 experiment, which has
searched for $e^+ e^- \to X \gamma$, followed by the decay $X \to e^+
e^-$~\cite{Anastasi:2015qla}, leading to 
\begin{align}\label{eq:4.6}
 {\varepsilon^V_{ee}}^2+{\varepsilon^A_{ee}}^2   < \frac{4 \times
   10^{-6}}{{\text{BR}(Z^\prime\to e^+e^-)}}\,. 
\end{align}
Similar searches were also performed at BaBar, although the latter
were only sensitive to slightly heavier candidates, with
masses $m_X > 20$~MeV~\cite{Lees:2014xha}.

\paragraph{Light meson decays}

\noindent
In addition to the (direct) requirements that an explanation of the $^8$Be anomaly
imposes on the couplings of the $Z^\prime$ to quarks - already discussed in
Section~\ref{sec:IPCcon}-, important constraints on the latter arise
from several light meson decay experiments. 
For instance, this is the case of 
searches for $\pi^0 \to \gamma Z^\prime(Z^\prime\to ee)$ and $K^+ \to \pi^+ Z^\prime(Z^\prime\to ee)$
at the NA48/2~\cite{Raggi:2015noa} experiment,
as well as searches for 
$\phi^+ \to \eta^+ Z^\prime(Z^\prime\to
ee)$ at KLOE-2~\cite{Anastasi:2015qla}.
Currently, the most stringent constraint does arise
from the rare pion decays searches which lead, for
$m_{Z^\prime} \simeq 17$ MeV~\cite{Raggi:2015noa}, to the following bound 
\begin{align}
\vert 2 \varepsilon^V_{uu}  + \varepsilon^V_{dd}  \vert \,=\,\vert
\varepsilon^V_{p}  \vert \,\lesssim \,\frac{1.2 \times
  10^{-3}}{\sqrt{\textrm{BR}(Z^\prime \to e^+ e^-)}}\, . 
\end{align}
If one confronts the range for $|\varepsilon^V_p + \varepsilon^V_n|$
required to explain the anomalous IPC in $^8$Be (see
Eq.~\eqref{eq:4.4}), with the comparatively small allowed regime for
$\vert \varepsilon^V_{p}  \vert$  from the above equation, it is
clear that in order to explain the anomaly in $^8$Be the neutron
coupling $\varepsilon^V_n$ must be sizeable 
(This enhancement of neutron couplings (or suppression of the proton
ones) is also often referred to as
a ``protophobic scenario'' in the literature). Further (subdominant)
bounds can also be obtained from 
neutron-lead scattering, proton fixed target experiments and other
hadron decays, but  we will not take them into account in the present study

\paragraph{Constraints arising from parity-violating experiments}

\noindent
Very important constraints on the product of vector and axial
couplings of the $Z^\prime$ to electrons arise from the parity-violating
M{\o}ller scattering,
measured at the SLAC E158 experiment~\cite{Anthony:2005pm}.
For $m_{Z^\prime} \simeq 17$ MeV, it yields~\cite{Kahn:2016vjr}
\begin{align}
\vert \varepsilon^V_{ee}  \varepsilon^A_{ee} \vert \lesssim 1.1 \times 10^{-7}.
\end{align}
Further useful constraints on a light $Z^\prime$ couplings can be
inferred from atomic parity violation in Caesium, in particular 
from the measurement of the effective weak charge of the
Cs atom~\cite{Davoudiasl:2012ag,Bouchiat:2004sp}. At the $2\sigma$
level~\cite{Porsev:2009pr}, these yield
\begin{align}\label{eqn:parity}
|\Delta Q_w|\, =\, |\frac{2\sqrt{2}}{G_F}\, 4\pi\alpha\,
\varepsilon^A_{ee} \left[\varepsilon^V_{uu} (2 Z + N) +
  \varepsilon^V_{dd} (Z + 2 N) \right] \left( \frac{\mathcal
  K(m_{Z^\prime})}{m_{Z^\prime}^2}\right)| \,\lesssim \,0.71\,, 
\end{align}
in which $\mathcal K$ is an atomic form factor, with
$\mathcal K(17 \,\mathrm{MeV})\simeq 0.8$~\cite{Bouchiat:2004sp}. 
For the anomalous IPC favoured values of $\varepsilon^V_{uu(dd)}$, 
the effective weak charge of the
Cs atom measurement provides a very strong constraint on
$|\varepsilon^A_{ee}|$, $|\varepsilon^A_{ee}| \lesssim 2.6 \times
10^{-9}$,
which is particularly relevant for our scenario, as it renders a
combined explanation of $(g-2)_e$ and the anomalous IPC particularly
challenging. As we will subsequently discuss, the constraints on
$|\varepsilon^A_{ee}|$ exclude a large region of the parameter space,
leading to a ``predictive'' scenario for the $Z^\prime$ couplings.

\paragraph{Neutrino-electron scattering experiments}

\noindent
Finally, neutrino--electron scattering provides stringent constraints
on the $Z^\prime$  neutrino
couplings~\cite{Bilmis:2015lja,Khan:2016uon,Lindner:2018kjo}, with the
tightest bounds arising from the TEXONO and CHARM-II experiments.
In particular, for the mass range $m_{Z^\prime} \simeq 17\:\mathrm{MeV}$,
the most stringent
bounds are in general due to the TEXONO
experiment~\cite{Deniz:2009mu}. While for some simple $Z^\prime$
constructions the couplings are flavour-universal, the extra fermion
content in our model leads to a decoupling of the lepton families in
such a way that only the couplings to electron neutrinos can be
constrained with the TEXONO data. 
For muon neutrinos, slightly weaker but nevertheless very
relevant bounds can be
obtained from the CHARM-II experiment~\cite{Vilain:1993kd}.

\subsection{Majorana neutrinos: fitting the leptonic axial and vector
  couplings}\label{sec:subsec:MajoranaFit} 

In the present model, neutrinos are Majorana particles, which implies
that the corresponding flavour conserving pure vector part of the
$Z^\prime$-couplings vanishes. The fits performed in
Refs.~\cite{Bilmis:2015lja,Khan:2016uon,Lindner:2018kjo} are thus not
directly applicable to our study; consequently we have performed
new two-dimensional fits to simultaneously constrain the axial
couplings to electron and muon neutrinos, and the vector coupling to
electrons,  following the prescription of Ref.~\cite{Lindner:2018kjo}.
As argued earlier, the axial coupling to electrons has to be
negligibly
small in order to comply with constraints from atomic parity violation
and, for practical purposes, these will henceforth be set to zero in our analyses.
 
\begin{figure}
\centering
\includegraphics[width=0.7\textwidth]{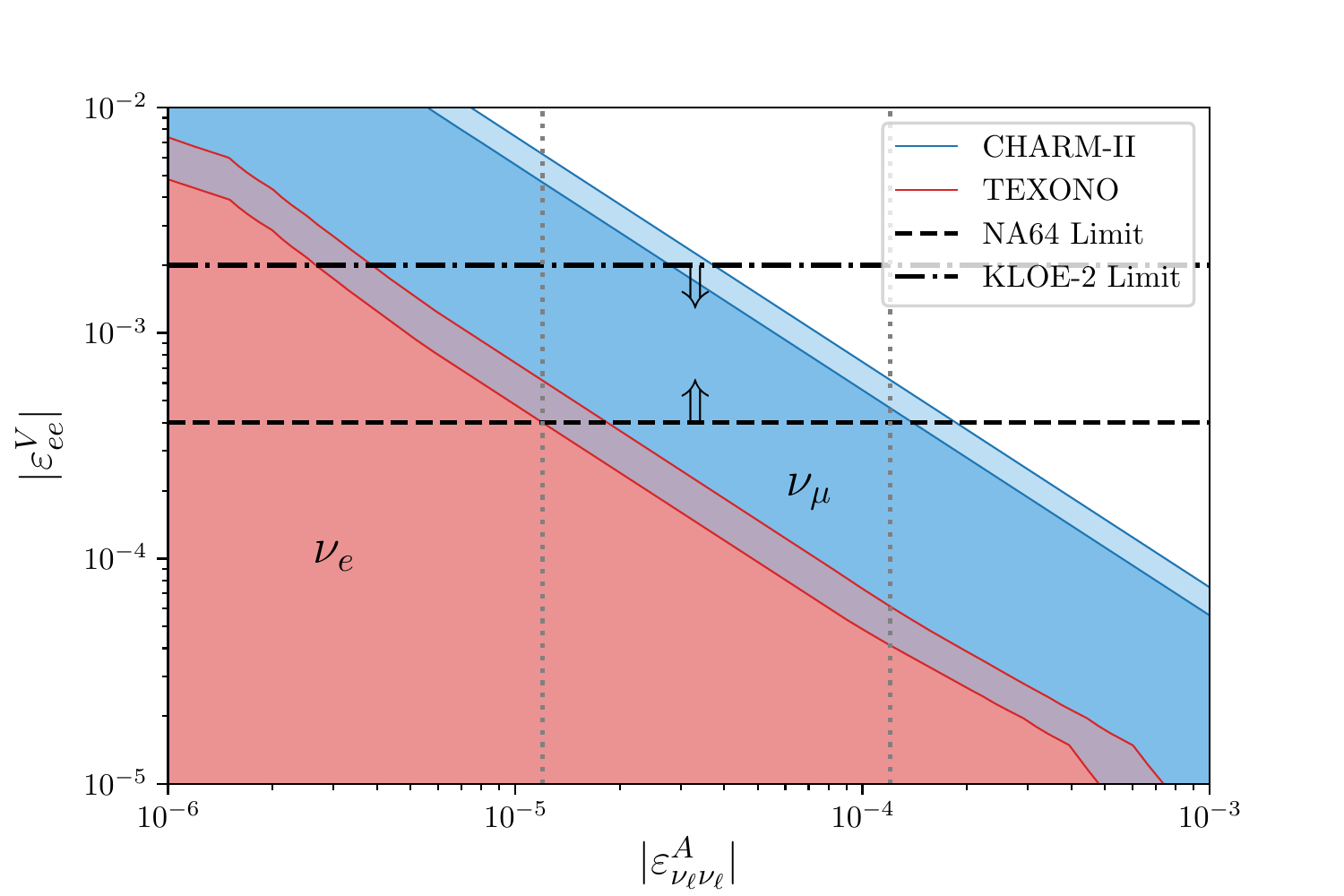}
\caption{New $\chi^2$-fit of the $\bar\nu_e\:e$ scattering data of TEXONO
  (red) and the $\bar\nu_\mu\:e$ scattering data of CHARM-II
  (blue), displaying the 1- and 2-$\sigma$ allowed regions around the best fit
  point (respectively darker and lighter colours).
 The lower bound of NA64 (dashed line) and
  the upper bound by KLOE-2 (dash-dotted line) are also shown, with
  the arrows identifying the viable allowed regions. The obtained upper
  limits on the axial coupling to neutrinos,
  cf. Eq~\eqref{eqn:nu_scat_lim}, are marked by dotted lines:
  the TEXONO data mostly constrains the couplings to electron neutrinos while
  the CHARM-II data is responsible for the constraints on 
  the couplings to muon neutrinos. 
}
\label{fig:nu_scat}
\end{figure}

In Fig.~\ref{fig:nu_scat} we show the particular likelihood contours
deviating $1$ and $2\,\sigma$ from the best fit point for the
neutrino--electron scattering data, which is found to lie very close
to the SM prediction. 
Applying the constraints on the electron vector coupling
$\varepsilon^V_{ee}$ obtained from NA64~\cite{Banerjee:2018vgk} and KLOE-2~\cite{Anastasi:2015qla} leads to the
limits
\begin{eqnarray}\label{eq:nuelim}
	|\varepsilon_{\nu_e\nu_e}^A| &\lesssim& 1.2\times
	10^{-5}\,\text,\nonumber\\ 
	|\varepsilon_{\nu_\mu\nu_\mu}^A| &\lesssim& 12.2\times 10^{-5}\,\text,
	\label{eqn:nu_scat_lim}
\end{eqnarray}
leading to which we have assumed
the smallest allowed electron coupling $|\varepsilon_{ee}^V| \sim
4\times 10^{-4}$.
Note that interference effects between the charged and neutral
currents (as discussed in
Refs.~\cite{Feng:2016ysn,Bilmis:2015lja,Khan:2016uon,Lindner:2018kjo})
do not play an important role in this scenario, due to vanishing
neutrino vector couplings. 
The technical details regarding the calculation and fitting procedure
referred to above can be found in the Appendix.

\bigskip
To conclude this section, we list below a summary of the relevant
constraints so far inferred on the couplings of the $Z^\prime$ to matter: combining the required ranges of couplings needed to explain the
anomalous IPC signal with the relevant bounds from other experiments,
we have established the following ranges for the couplings (assuming
$\text{BR}(Z^\prime \to e^+ e^-) = 1$),
\begin{eqnarray}
2\times 10^{-3} \;\lesssim\;| \varepsilon^V_n | &\;\lesssim\;& 15
\times 10^{-3}\, , 
\\
| \varepsilon^V_p | &\;\lesssim \; & 1.2 \times 10^{-3}\, ,
\\
0.4\times 10^{-3} \;\lesssim\;| \varepsilon^V_{ee} | &\;\lesssim\;&  2
\times 10^{-3}\, , 
\label{eq:lepton:allowed:couplings1}
\\
| \varepsilon^A_{ee} | &\;\lesssim\;& 2.6 \times 10^{-9}\, ,\\
|\varepsilon^A_{\nu_e\nu_e}| &\lesssim& 1.2\times 10^{-5}\, ,\\
|\varepsilon^A_{\nu_\mu\nu_\mu}| &\lesssim& 12.2\times 10^{-5}\, .
\end{eqnarray}
%


\section{Addressing the anomalous IPC in $\pmb{^8}$Be: impact for a combined
explanation of $\pmb{(g-2)_{e,\mu}}$}\label{sec:combined:explanation}

As a first step, we apply the previously obtained model-independent
constraints on the $Z^\prime$ couplings to the specific structure of
the present model. After taking the results of (negative) collider
searches for the exotic matter fields into account, we will be able to infer
an extremely tight range for $\varepsilon$ (which we recall to
correspond to a redefinition of the effective kinetic mixing
parameter, cf. Eq.~(\ref{eq:epsilon:redefine})). In turn, this will
imply that very little freedom is left to explain the experimental
discrepancies in the light charged lepton anomalous
magnetic moments, the latter requiring an interplay of the $h_{\ell \ell}$
and $k_{\ell \ell}$ couplings.

\subsection{Constraining the model's 
parameters}\label{subsec:model-constraint}
The primary requirements  to 
explain the anomalous IPC in $^8$Be concern 
the physical mass of the $Z^\prime$, which should approximately be  
\begin{equation}  \label{eq:7.1}
  m_{Z^\prime} \,\approx 17\,\mathrm{MeV}\,,
\end{equation}
and the strength of its couplings to nucleons (protons and neutrons),
as given in Eqs.~(\ref{eqn:epsn}, \ref{eqn:epsp}). With 
$\varepsilon^V_{qq}$ as defined in Eq.~\eqref{eq:epsq},
and recalling that
$\varepsilon_p = 2\,\varepsilon^V_{uu} + \varepsilon^V_{dd}$ and
$\varepsilon_n = \varepsilon^V_{uu} + 2 \,\varepsilon^V_{dd}$, one
obtains the following constraints on $\varepsilon_{B-L}$ and
$\varepsilon$
\begin{eqnarray}
  |\varepsilon_n^V| &=& |\varepsilon_{B-L}| \,=\,
  (2 - 15)\times 10^{-3}\,
  , \label{eq:7.2}\\ 
  |\varepsilon_p^V| &=& |\varepsilon + \varepsilon_{B-L}| \lesssim 1.2
  \times 10^{-3}\,. 
  \label{eq:7.3}
\end{eqnarray}
Furthermore, this implies an upper bound for the VEV of $h_X$,
$v_X \lesssim 14\:\mathrm{GeV}$, since
\begin{equation}
  m_{Z^\prime} \approx m_{B^\prime} \,=\, 2\, e\, \vert
  \varepsilon_{B-L}\vert \,
  v_X\,.
\end{equation}

In the absence of heavy vector-like leptons, there are no other
sources of mixing in the lepton section in addition to the PMNS.
This would imply that the effective couplings of the $Z^\prime$ to
neutrinos are identical to that of the neutron (up to a global sign),
that is
\begin{equation}  
\label{eq:7.4}
  \varepsilon_{\nu\nu}^A \, = \, \varepsilon_{B-L}\,, 
\end{equation}
which, in view of Eq.~\eqref{eq:7.2}, leads to
$ \varepsilon_{\nu\nu}^A = (2 - 10)\times 10^{-3}$. However,
the bounds of the TEXONO experiment~\cite{Deniz:2009mu} for 
neutrino-electron scattering (cf. Eq.~\eqref{eq:nuelim})
imply that for the minimal allowed electron coupling
$|\varepsilon_{ee}^V| \gtrsim 0.4\times 10^{-3}$ one requires 
\begin{equation}  \label{eq:7.5}
  |\varepsilon_{\nu\nu}^A|\,\lesssim\, 1.2 \,(12.2)\,\times\, 10^{-5}\,,
\end{equation}
for electron (muon) neutrinos. As can be inferred, this is 
in clear conflict with the values of $ \varepsilon_{\nu\nu}^A$
required to explain the $^8$Be anomalous IPC, which are $\mathcal{O}(10^{-3})$. 

\noindent
In order to circumvent this problem,
the effective $Z^\prime$ coupling to the SM-like neutrinos must be
suppressed. The additional vector-like leptons open the possibility
of having new sources of mixing between the distinct species of
neutral leptons; the effective neutrino coupling derived in 
Section~\ref{section:newneutralcurrent}
allows to suppress the couplings by a factor
$\sim (1 - {\lambda_{L\,\alpha}^2 v_{X}^2/M_{L\,\alpha}^2})$ (see
Eq.~(\ref{eqn:nunu}), with $\alpha$ denoting SM flavours), hence implying
\begin{equation}  \label{eq:7.6}
  |1 - \frac{\lambda_L^2 v_X^2}{M_L^2}| \,\lesssim 0.01\,.
\end{equation}
Thus, up to a very good approximation, we can assume 
$\lambda_L v_X\simeq M_L$ for each lepton generation $\alpha$.   
On the other hand, from Eqs.~\eqref{eq:epsl} and~\eqref{eqn:parity} it
follows that the bound from atomic parity violation in Caesium tightly
constrains the isosinglet vector-like lepton coupling
$\lambda_E$ (for the first lepton generation)\footnote{In what
  follows, we will not explicitly include the flavour indices, as it
  would render the notation too cumbersome, but rather describe it in
  the text.}, leading to 
\begin{eqnarray}
  |\varepsilon_{ee}^A| \,=  \,\left|\frac{1}{2}\left(\frac{\lambda_E^2 \,
    v_X^2}{M_E^2} - \frac{\lambda_L^2 \,
    v_X^2}{M_L^2}\right)\varepsilon_{B-L}\right| \,\lesssim \,
  2.6\times10^{-9}\,\text, 
  \label{eqn:limit_axial_0}
\end{eqnarray}
which in turn implies 
\begin{eqnarray}
  \left|\frac{\lambda_E^2  \,v_X^2}{M_E^2}
  - \frac{\lambda_L^2  \,v_X^2}{M_L^2}\right|
   \,\lesssim 2.6\times 10^{-6}\,\text.
  \label{eqn:limit_axial}
\end{eqnarray}
Notice that this leads to a tight correlation between the
isosinglet and isodoublet vector-like lepton couplings,
$\lambda_E$ and $\lambda_L$, respectively.
More importantly, the above discussion renders manifest the
necessity of having the additional field content (a minimum of two
generations of heavy vector-like leptons).

Together with Eqs.~\eqref{eqn:eeL} and~\eqref{eqn:eeR},
Eqs.~\eqref{eq:7.6} and~\eqref{eqn:limit_axial}
suggest that the $Z^\prime$ coupling to electrons is now almost solely
determined by $\varepsilon$.
In particular, the KLOE-2~\cite{Anastasi:2015qla} limit of Eq.~\eqref{eq:4.6} for
$\varepsilon_{ee}$ now implies
\begin{eqnarray}
  |\varepsilon|\,< \,0.002\,.
  \label{eqn:limit_eps}
\end{eqnarray}

Further important constraints on the model's parameters arise from the 
masses of the vector-like leptons, which are bounded from both below
and above. On the one hand, 
the perturbativity limit of the couplings $\lambda_L$ and $\lambda_E$
implies an upper bound on the vector-like lepton masses.
On the other hand, direct searches for vector-like leptons exclude
vector-like lepton masses below
$\sim 100\,\mathrm{GeV}$~\cite{Achard:2001qw}
(under the assumption these dominantly decay into $W\nu$ pairs).
This bound can be relaxed if other decay modes exist, for instance
involving the $Z^\prime$ and $h_X$ as is the case in our scenario.
However, and given the similar decay and production mechanisms,
a more interesting possibility is to recast the results of LHC
dedicated searches for sleptons
(decaying into a neutralino and a charged SM lepton) for the case of
vector-like leptons decaying into $h_X$ and a charged SM lepton. 
Taking into account
the fact that the vector-like lepton's cross section is
a few times larger than the selectron's or
smuon's~\cite{Feng:2016ysn,Khachatryan:2014qwa}, one can roughly
estimate that vector-like leptons with a mass $\sim 100\,\mathrm{GeV}$
can decay into a charged lepton and an $h_X$ with mass
$\sim~(50-70)\,\mathrm{GeV}$. 
Therefore, as a benchmark choice we fix the tree-level mass of the
vector-like leptons of all generations to $M_L = M_E = 90\,\mathrm{GeV}$
(which yields a physical mass $\sim110\,\mathrm{GeV}$, once the
corrections due to mixing effects are taken into account). In turn,
this implies that the couplings $\lambda_{L,E}$ should be sizeable 
$\lambda_E^e \approx \lambda_L^e \sim 6.4$ (for the first
generation, due to the very stringent parity violation
constraints)\footnote{Couplings so close to the perturbativity limit of
  $\mathcal{O}(4\pi)$ can potentially lead to Landau poles at
  high-energies, as a consequence of running effects. To avoid this,
  the low-scale model here studied should be embedded into an
  ultra-violet complete framework.}, while for the second generation
one only has $\lambda_L^\mu \sim 6.4$. 
(We notice that
smaller couplings, still complying
with all imposed constraints can still be accommodated, at the price
of extending the particle content to include additional exotic fermion
states.)
In agreement with the the above discussion, we further 
choose $m_{h_X} = 70 \,\mathrm{GeV}$ as a benchmark value. Since
$h_X$ can also decay into two right handed neutrinos (modulo a
substantially large Majorana coupling $y_M$), leading to a signature
strongly resembling that of slepton pair production, current negative
search results then lead to constraints on $\varepsilon_{B-L}$.
For the choice $m_{h_X} = 70 \,\mathrm{GeV}$, $\varepsilon_{B-L}$
should be close to its smallest allowed value $\varepsilon_{B-L} =
0.002$ ~\cite{Feng:2016ysn}, which in turn implies the following range
for $\varepsilon$
 \begin{eqnarray}
  -0.0032 \,\lesssim \,\varepsilon\, \lesssim\, -0.0008\,.
   \label{eqn:limit_eps_coll}
 \end{eqnarray}
The combination of the previous constraint with the one inferred from
the KLOE-2 limit on the couplings of the $Z^\prime$ to electrons,
see Eq.~\eqref{eqn:limit_eps}, allows to derive the viability range for
$\varepsilon$,  
\begin{eqnarray}
 -0.002 \,\lesssim \,\varepsilon \,\lesssim \,-0.0008\,.
  \label{eqn:limit_eps_coll2}
\end{eqnarray}

\bigskip

Before finally addressing the feasibility of a combined explanation to
the atomic $^8$Be and $(g-2)_{e,\mu}$ anomalies, let us notice
that in the study of
Ref.~\cite{Dror:2017nsg} the authors have derived significantly stronger
new constraints on the parameter space of new (light) vector states,
$X$, arising in $U(1)_X$ extensions of the SM, such as  $U(1)_{B-L}$
models. The new bounds can potentially disfavour some well-motivated
constructions, among which some aiming at addressing the $^8$Be
anomalies, and arise in general from an energy-enhanced emission
(production) of the longitudinal component ($X_L$) via anomalous
couplings\footnote{As discussed in~\cite{Dror:2017nsg}, 
such an enhancement can occur if the model's content is
such that a new set of heavy fermions with vector-like couplings to
the SM gauge bosons, but chiral couplings to $X$, is introduced to
cancel potentially dangerous chiral anomalies. Explicit Wess-Zumino
terms must be introduced to reinforce the SM gauge symmetry, which in
turn breaks the $U(1)_X$, leading to an energy-enhanced emission of
$X_L$. Moreover, the SM current that $X$ couples to may also be broken
at tree level, due to weak-isospin violation ($W \bar{u} d$ or 
$W\ell\bar{\nu}$ vertices may break $U (1)_X$, if $X$ has different
couplings to fermions belonging to a given $SU(2)_L$ doublet and
lacks the compensating coupling to the $W$).
In such a situation the
longitudinal $X$ radiation from charged current processes can be again
enhanced, leading to very tight constraints from 
$\pi\rightarrow e\nu_e +X$, or $W \to \ell \nu_\ell + X$. }. 
We notice that the prototype model here investigated departs in
several points from the assumptions of~\cite{Dror:2017nsg}.  In the
present  $U(1)_{B-L}$ extension the heavy vector-like fermions do not
contribute to anomaly cancellation, as the SM field content and three
generations of right handed neutrinos (and independently, all the
vector-like fermions) constitute a completely anomaly-free set under
the extended gauge group. 
Moreover, there are no neutral vertices explicitly breaking 
$SU(2)_L$ -- as can be seen from Eqs.~\eqref{eqn:nunu} and \eqref{eqn:eeL}, thus avoiding the
potential constraints inferred for a possible energy-enhanced longitudinal emission of $X$.

\subsection{A combined explanation of $\pmb{(g-2)_{e,\mu}}$}
In view of the stringent constraints on the parameter space of the
model, imposed both from phenomenological arguments and from a
satisfactory explanation of the anomalous IPC in $^8$Be, one must now
consider whether it is still possible to account for the observed
tensions in the electron and muon anomalous magnetic moments. As
discussed both in the Introduction and in Section~\ref{sec:g-2}, the
discrepancies between SM prediction and experimental observation have
an opposite sign for electrons and muons, and exhibit a scaling
behaviour very different from the na\"ive expectation (powers of the
lepton mass ratio). 

Given the necessarily small mass of the $Z^\prime$ and the large couplings
between SM leptons and the heavier vector-like states
($\lambda_{L,E}$), in most of the parameter space
the new contributions to $(g-2)_{e,\mu}$ are
considerably larger than what is suggested from experimental
data. Firstly, recall that due to the opposite sign of the loop
functions for (axial) vector and 
(pseudo)scalar contributions, a cancellation between the latter
contributions allows for an overall suppression of each $(g-2)_{e,\mu}$. 
Moreover, a partial cancellation between the distinct diagrams can
lead to $\Delta a_\mu$ and $\Delta a_e$ with opposite signs; this
requires nevertheless a large axial coupling to electrons, which is 
experimentally excluded. However, an asymmetry in the
couplings of the SM charged leptons to the vector-like states
belonging to the same generation can overcome the problem,
generating a sizeable ``effective'' axial coefficient
$g_A^{\ell\ell}$: while for electrons Eq.~\eqref{eqn:limit_axial}
implies a strong relation between $\lambda_L$ and $\lambda_E$, the
(small) couplings $h_{\ell}$ and $k_{\ell}$ remain essentially 
unconstrained\footnote{Being diagonal in generation space, we
  henceforth denote the couplings via a single index, i.e.
  $h_{\ell} = h_{\ell\ell}$, etc., for simplicity.} and can induce
such an asymmetry, indeed leading to the desired ranges for the
anomalous magnetic moments. 

\begin{figure}
    \begin{center}
    \mbox{\hspace{-1cm}
      \includegraphics[width=0.55\textwidth]{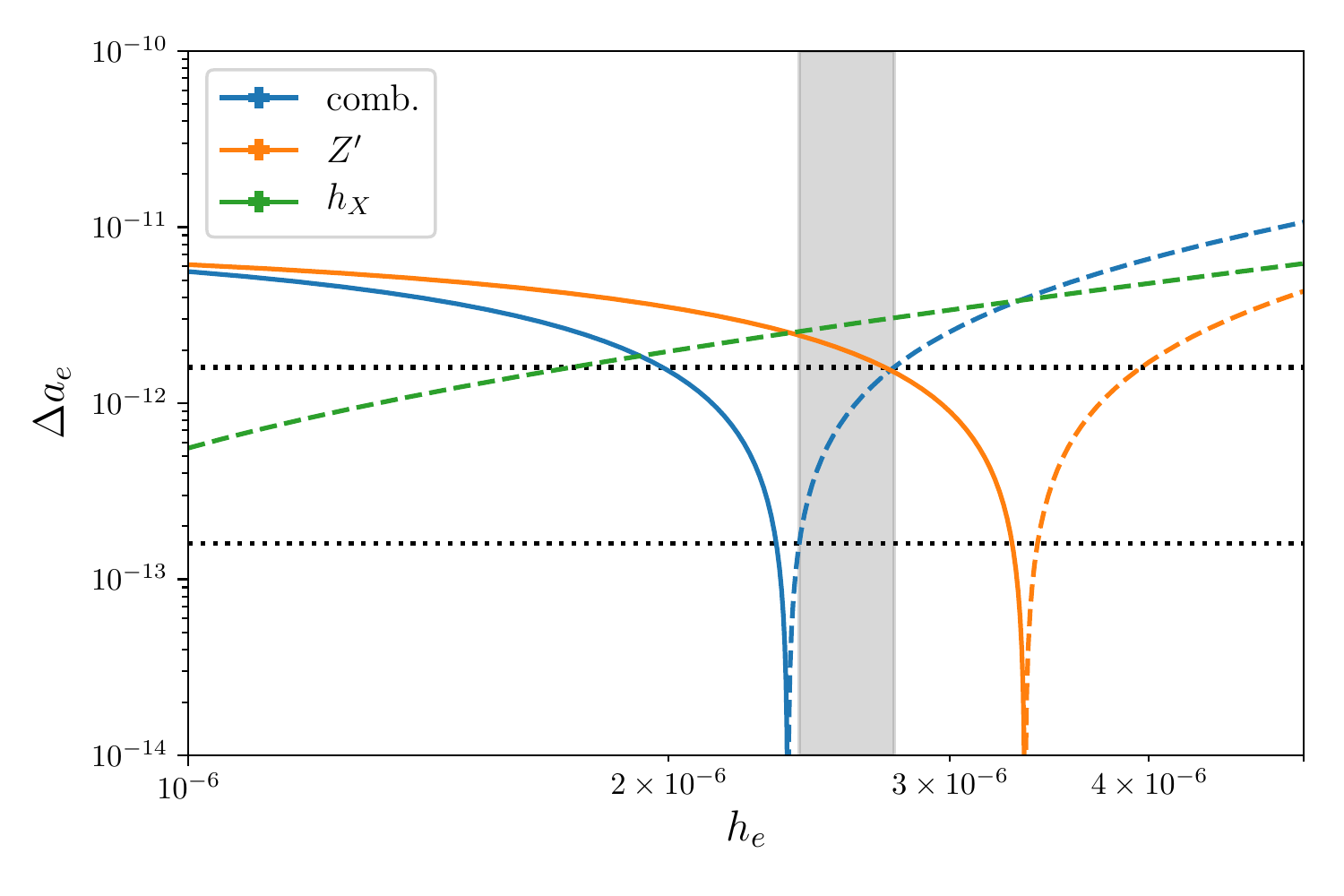} 
   	\includegraphics[ width=0.55\textwidth]{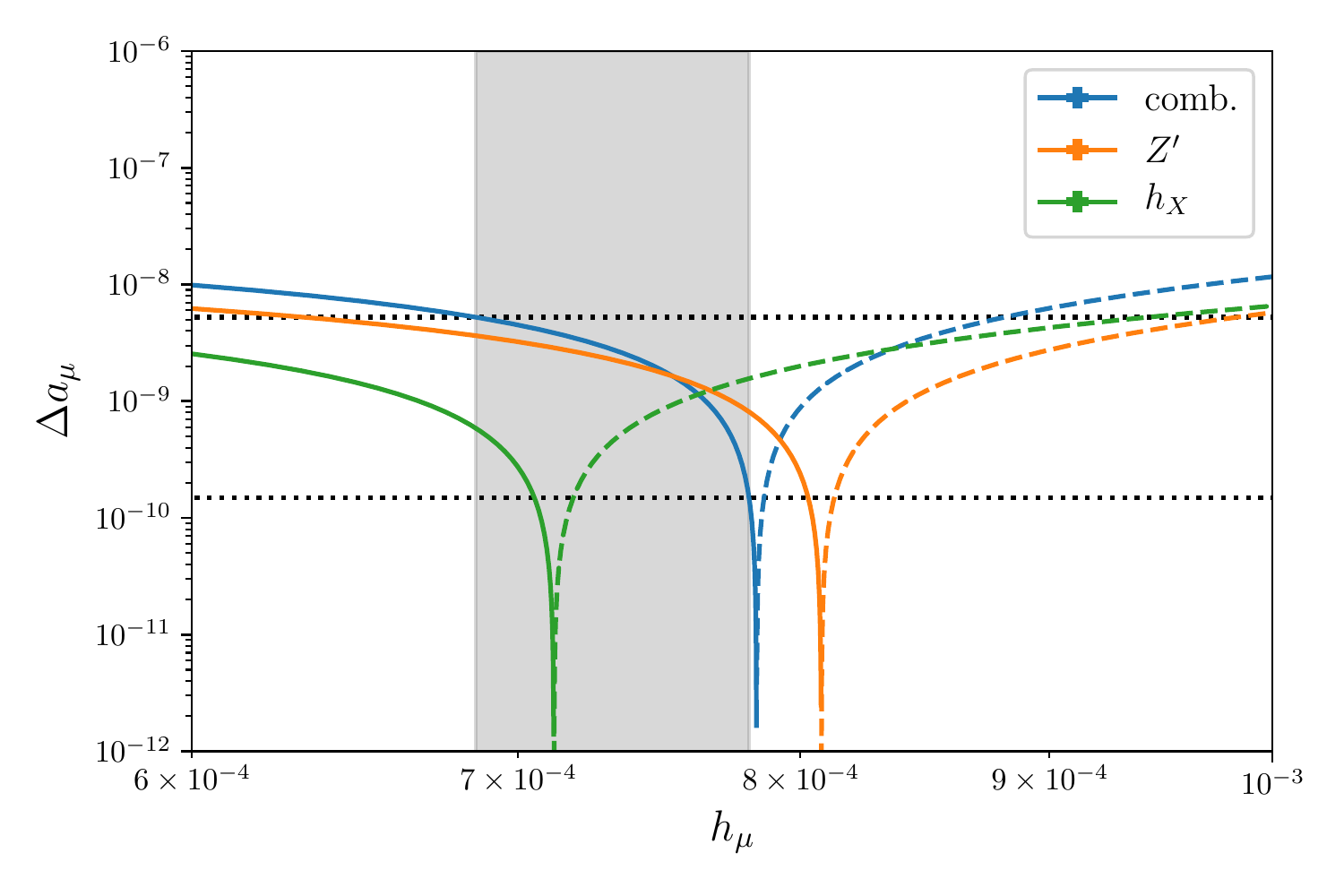}}
    \end{center}
    \caption{Contributions to the anomalous magnetic moment of charged
      leptons, $|\Delta a_\ell|$, as a function of the $h_{\ell}$
      coupling for $\ell=e$ (left) and  $\ell=\mu$ (right).
      Solid (dashed) lines correspond to positive (negative) values of 
      $\Delta a_\ell$; the colour code denotes contributions from the
      $Z^\prime$ (orange) and from $h_X$ (green), as well as the
      combined one (blue). Horizontal (dotted) lines denote the
      $2\sigma$ and $3\sigma$ regions of the electron and muon
      $\Delta a_\ell$. A vertical opaque region corresponds to the
      $h_{\ell}$ interval for which the
      combined contributions to $\Delta a_{e (\mu)}$
      lie within the $2\sigma$ ($3\sigma$) region.  
      Leading to this figure, we have selected a benchmark choice of
      parameters complying with all the constraints mentioned in
      Section~\ref{sec:phenocon}: $M_L = M_E = 90\,\mathrm{GeV}$,
      $\lambda_E = \lambda_L = M_L / v_X$, $m_{h_X} = 70\,\mathrm{GeV}$,
      $\varepsilon = -8\times10^{-4}$, $\varepsilon_{B-L} = 0.002$ and 
      $k_\ell = 10^{-7}$.}
\label{fig:cancellation}
\end{figure}

\begin{figure}[t!]
  \begin{center}
    \mbox{
 \hspace{-1.5cm}\includegraphics[width = 0.65 \textwidth]{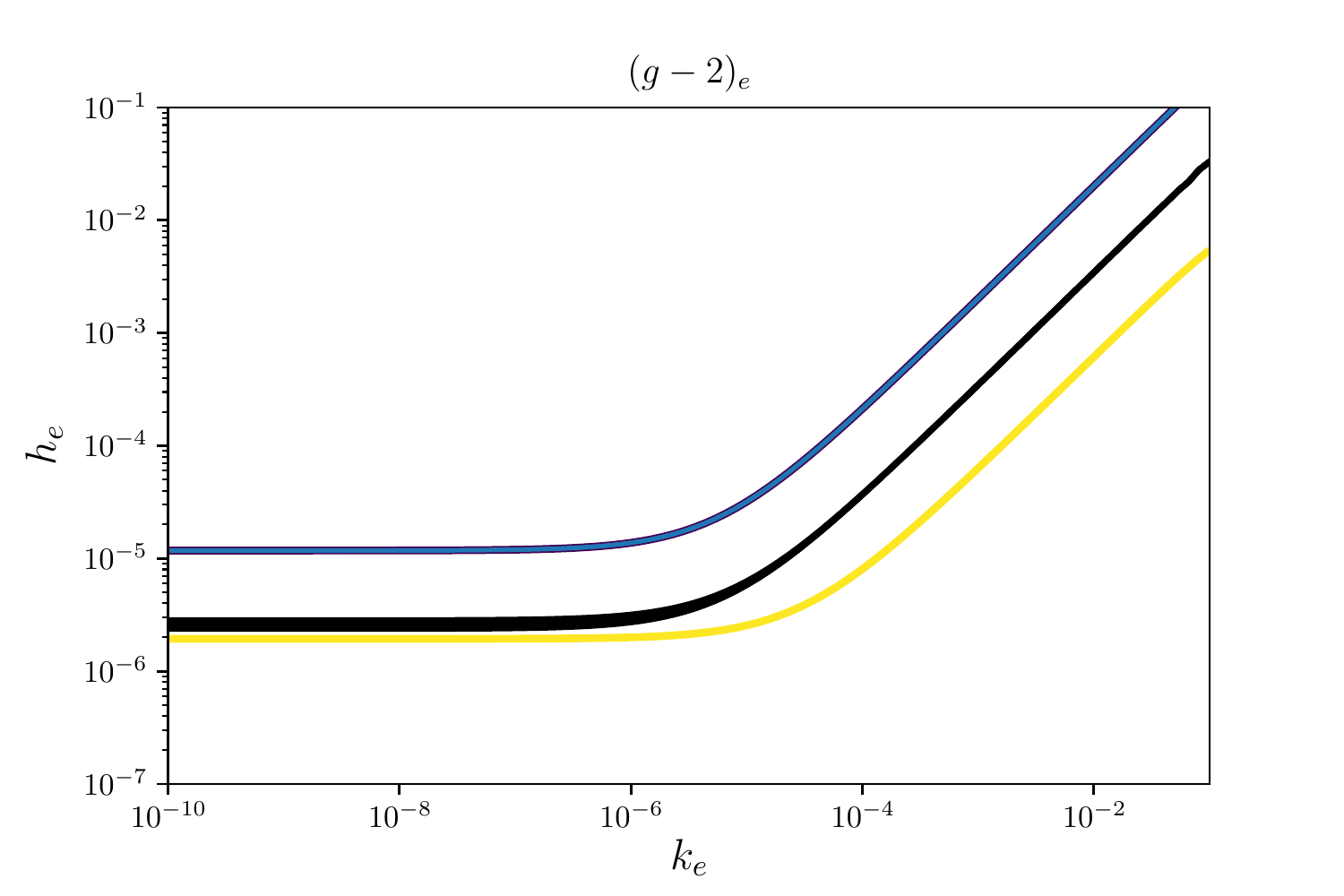}
 \hspace{-1.2cm}\includegraphics[width = 0.65 \textwidth]{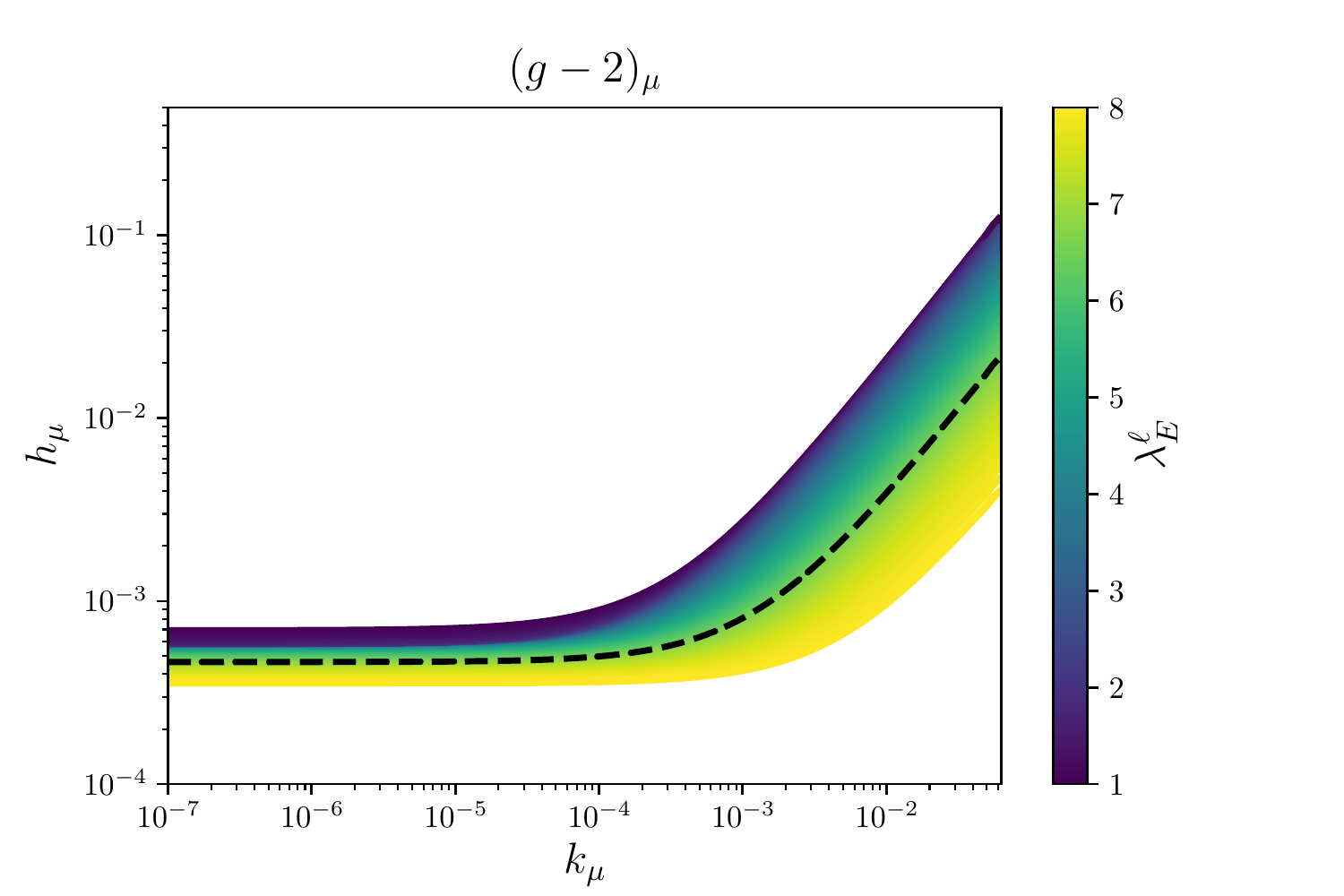}} 
\caption{Viable regions in $h_\ell$ vs. $k_\ell$ parameter space: on the left
  (right) $\ell=e\, (\mu)$). In both panels the colour code denotes
  the value of $\lambda_E^\ell$ ($\lambda_E = 1- 8$, from dark violet
  to yellow). 
  On the left panel, only the central black line complies with
  $(g-2)_e$ at the $2 \sigma$ level (i.e. $\lambda_E^e\sim  6.4$);
  for the right panel, all the coloured region allows to satisfy
   $(g-2)_\mu$ at $2 \sigma$ (the dashed black line illustrates the
  value $\lambda_E^\mu\sim  6.4$). All other relevant parameters
  fixed as leading to Fig.~\ref{fig:cancellation}.
    }
    \label{fig:g-2comb}
  \end{center}
\end{figure}

This interplay of the different (new) contributions can be understood
from Fig.~\ref{fig:cancellation}, which illustrates the $h_X$ and the
$Z^\prime$ contributions to the electron and muon $|\Delta a_\ell|$, as a
function of the $h_{\ell}$ coupling for $\ell=e$ (left) and 
$\ell=\mu$ (right).
The $h_X$-induced contribution to $(g-2)_\ell$
changes sign when the pseudoscalar dominates over the scalar
contribution (for the choices of the relevant Yukawa couplings
$h_\ell$ and $k_\ell$). Likewise, 
a similar effect occurs for the $Z^\prime$ contribution
when the axial-vector contribution dominates over the vector one.
The transition
between positive (solid line) and negative (dashed line) contributions
- from $Z^\prime$ (orange), $h_X$ (green) and combined (blue) -
is illustrated by the sharp kinks visible in the logarithmic plots
of Fig.~\ref{fig:cancellation}.
In particular, notice that the negative electron $\Delta a_e$ is
successfully induced by the flip of the sign of the $h_X$
contribution, while a small positive muon $\Delta a_\mu$ arises from
the cancellation of the scalar  and the $Z^\prime$ contributions.
Leading to the numerical results of  Fig.~\ref{fig:cancellation}
(and in the remaining of our numerical analysis), we have taken as 
benchmark values $\varepsilon_{B-L} = 2\times 10^{-3}$ and
$\varepsilon = -8\times 10^{-4}$
(which are consistent with the criterion for explaining
the anomalous IPC in $^8$Be and respect all other imposed constraints). 
We emphasise that as a consequence of their already extremely
constrained ranges, both the $B-L$ gauge coupling and the kinetic
mixing parameter have a very minor influence on the contributions
to the anomalous magnetic lepton moments (when varied in the allowed
ranges). 
Furthermore, the masses $M_{L,E}$ and $m_{h_X}$ can be slightly varied 
with respect to the proposed benchmark values, with only a minor impact 
on the results; a mass-splitting between $M_E$ and $M_L$ (for each generation) slightly modifies the slope of the curves presented in Fig.~\ref{fig:g-2comb}, while an overall scaling to increase $M_{L,E}$ 
would imply taking (even) larger values for most of the couplings in their allowed regions. (Notice however that the model's parameter space is severely constrained, so that any departure from the benchmark values is only viable for a comparatively narrow band in the parameter space.)

To conclude the discussion, and provide a final illustration of how
constrained the parameter space of this simple model becomes, we
display in Fig.~\ref{fig:g-2comb}
the regions complying at the $2 \sigma$ level 
with the observation of $(g-2)_\ell$
in the planes spanned by $h_\ell$ and $k_\ell$ 
(for $\ell=e,\mu$). The colour code reflects the size of the
corresponding entry of $\lambda_E^\ell$, which
is varied in the interval $[1,8]$ (recall that for the electron
anomalous magnetic moment, $\lambda_L^e=\lambda_E^e\sim  6.4$). 
All remaining parameters are fixed to the same values used for the numerical 
analysis leading to Fig.~\ref{fig:cancellation}.

Notice that, as mentioned in the discussion at the beginning of the
section (cf. \ref{subsec:model-constraint}), the extremely stringent
constraints on the $Z^\prime$ couplings arising from  atomic parity
violation and electron neutrino scatterings render the model
essentially predictive in what concerns $(g-2)_e$: only the narrow
black band of the $(h_e-k_e)$ space succeeds in complying with all
available constraints, while both addressing the IPC $^8$Be anomaly, and
saturating the current discrepancy between SM and observation on  
$(g-2)_e$. For the muons, and although $h_\mu$ remains strongly
correlated with $k_\mu$, the comparatively larger freedom associated
with $\lambda_E^\mu$ (recall that no particular relation between
$\lambda_L$ and $\lambda_E$ is required by experimental data) 
allows to identify a wider band in $(h_\mu-k_\mu)$
space for which $\Delta a_\mu$ is satisfied at  $2 \sigma$. 

Finally, notice that the $h_\ell$ and  $k_\ell$ are forced into
a strongly hierarchical pattern, at least in what concerns the first
two generations.

\section{Concluding remarks}\label{sec:concs}
Despite the absence of a direct discovery of new resonances at high
energy colliders, several low-energy observables exhibit tensions with
SM predictions, to various degrees of significance and longevity. 
The discrepancy between the SM prediction and experimental observation
regarding the anomalous magnetic moment of the muon is perhaps the
most longstanding anomaly, currently exhibiting a tension around 
$3.3\sigma$; more recently, the electron $(g-2)$ also started to
display tensions between theory and observation (around $2.5\sigma$), 
all the most intriguing since instead of following a na\"ive scaling
proportional to powers of the light lepton masses, the comparison of 
$\Delta a_{e,\mu}$ suggests the presence of a New Physics which
violates lepton flavour universality. In recent years, an anomalous
angular correlation was observed for the 18.15~MeV nuclear transition of 
$^8$Be atoms, in particular an enhancement of the IPC at large angles,
with a similar anomaly having been observed in  $^4$He transitions. 

An interesting possibility is to interpret the atomic
anomalies as being due to the presence of a light vector boson, with a
mass close to $17$~MeV. Should such a state have non-vanishing
electroweak couplings to the standard fields, it could also have an
impact on $\Delta a_{e,\mu}$. 
In this work, we have investigated the phenomenological implications
of a BSM construction in which the light vector boson arises from a
minimal extension of the gauge group via an additional
$U(1)_{B-L}$. Other than the scalar field (whose VEV is responsible
for breaking the new $U(1)$), three generations of  Majorana right-handed
neutrinos, as well as of heavy vector-like leptons are added to the SM
field content. As discussed here, the new matter fields play an
instrumental role both in providing additional sources of leptonic
mixing, and in circumventing the very stringent experimental constraints. 

After having computed the modified couplings, and summarised the
model's contributions to the light charged lepton anomalous magnetic
moments, we identified how addressing the anomalous IPC in $^8$Be
constrained the couplings of matter to the new $Z^\prime$.  
Once all remaining phenomenological constraints are imposed on the
model's parameter space, one is led to an extremely tight scenario,
in which saturating the (opposite-sign) 
tensions on $\Delta a_{e,\mu}$ can only be achieved via a cancellation
of the new (pseudo)scalar and (axial)vector contributions. 

The very stringent constraints arising from atomic parity violation
lead to an extremely strong correlation for the new couplings $h_e$ and
$k_e$ (first generation couplings of vector-like leptons to 
the SM Higgs) in order to comply with $(g-2)_e$; moreover, this
requires a nearly 
non-perturbative regime for the $\ell-L-h_X$ couplings
($\lambda_{L,E}$). The situation is slightly less constraining for the
muon anomalous magnetic moment, albeit leading to a non-negligible
dependence of the corresponding couplings, $h_\mu$ and $k_\mu$.

\medskip
Future measurements of 
right-handed neutral couplings, or axial couplings, for the second
generation charged leptons could further constrain the new muon
couplings. Although this clearly goes beyond the scope of the present work,
one could possibly envisage parity-violation experiments carried in
association with muonic atoms.
As an example, in experiments designed to test parity non-conservation (PNC) with atomic radiative capture (ARC), the 
measurement of the forward-backward asymmetry of the photon 
radiated by muons ($2s \to 1s$ transition) is sensitive to 
(neutral) muon axial couplings~\cite{McKeen:2012sh}. Further 
possibilities include scattering experiments, such as MUSE at 
PSI~\cite{Gilman:2017hdr}, or  
studying the muon polarisation in $\eta$ decays (REDTOP 
experiment proposal~\cite{Gatto:2019dhj}), which could allow a measurement of the 
axial couplings of muons.

\section*{Acknowledgements}
CH acknowledges support from the DFG Emmy Noether Grant No. HA
8555/1-1. JK, JO and AMT acknowledge support within the framework of
the European Union's Horizon 2020 research and innovation programme
under the Marie Sklodowska-Curie grant agreements No 690575 and No
674896. 


\begin{appendix}

\section*{Appendix}

\section*{Electron-neutrino scattering data: analysis and fits}
\label{sec:nu_scat}
In this Appendix we detail the formalism 
used for the analysis of the
available data on electron-neutrino scattering, 
upon which rely the constraints presented in 
Section~\ref{sec:phenocon:Zprime}. The data used arises from two
different experimental set-ups, CHARM-II and TEXONO. 

In general, the differential cross section for neutrino and
antineutrino scattering can be easily computed~\cite{Lindner:2018kjo}
\begin{eqnarray}
\frac{d\sigma}{dT}(\bar\nu e^- \to \bar\nu e^-) &=
\frac{m_e}{4\,\pi}\left[G_+^2 + G_-^2\left(1 - \frac{T}{E_\nu} \right)^2
  - G_+ G_- \frac{m_e \,T}{E_\nu^2} \right]\,\text,\\ 
\frac{d\sigma}{dT}(\nu e^- \to \nu e^-) &= \frac{m_e}{4\,\pi}\left[G_-^2
  + G_+^2\left(1 - \frac{T}{E_\nu} \right)^2 - G_+ G_- \frac{m_e\,
    T}{E_\nu^2} \right]\,\text, 
\end{eqnarray}
where $T$ is the recoil energy of the electron and $E_\nu$ the energy
of the (anti)neutrino. 
The coefficients $G_{\pm}$ are defined as
\begin{eqnarray}
G_{\pm} \,=\, \sum_{i = W, Z, Z^\prime} \frac{1}{P_i}\,
(g_{V_i}^{\nu\nu} - g_{A_i}^{\nu\nu})\,(g_{V_i}^{ee} \pm g_{A_i}^{ee})\,\text.
\end{eqnarray}
In the above, the sum runs over all relevant vector bosons (i.e.
$W$, $Z$ and $Z^\prime$), with $P_i$ denoting the denominator 
of the corresponding propagators;  $g_{V_i}$ and $g_{A_i}$ correspond
to the vector and axial couplings of the involved vector bosons to
(anti)neutrinos and electrons.
Since the energy of the neutrinos is well below the masses 
of the relevant gauge bosons, we carry the following approximations
\begin{eqnarray}
P_W \approx -\frac{\sqrt{2}\,g^2}{8\,G_F}\,\text,\quad 
P_Z \approx -\frac{\sqrt{2}\,g^2}{8 \,G_F \,c_w^2}\,\text,\quad 
P_{Z^\prime} \sim -(2 \,m_e \,T +m_{Z^\prime}^2)\,\text. 
\end{eqnarray}
For the case of the model under study, 
the vector and axial coefficients are given by
\begin{eqnarray}
	g_{V_W} &=& - g_{A_W} = \frac{g}{2\,\sqrt{2}}\quad\text{(for
          both $\nu$ and $e$),}\\ 
	g_{A_Z}^{\nu\nu} &=& -\frac{g}{2\,c_w}\,\text,\\
	g_{V_Z}^{ee} &=& - \frac{g\,(1 - 4\,s_w^2)}{4\,c_w}\,\text,\\
	g_{A_Z}^{ee} &=& \frac{g}{4\,c_w}\,\text,\\
	g_{V_{Z^\prime}}^{ee} &=& e \varepsilon_{ee}^V\,\text,\\
	g_{A_{Z^\prime}}^{\nu\nu} &=&2\, e \,\varepsilon_{\nu\nu}^A\,\text,
\end{eqnarray}
with all other remaining coefficients vanishing.
In order to take into account the fact that for Majorana neutrinos
the $\nu$ and $\bar\nu$ final  states are indistinguishable, 
a factor of 2 is present in the (axial) neutrino coefficients
(effectively allowing to double the contributions from amplitudes
involving two neutrino operators~\cite{Rosen:1982pj}).

\medskip
\paragraph{Data from the CHARM-II experiment}

\noindent
To fit the data from the CHARM-II experiment (extracted from 
Table 2 of Ref.~\cite{Vilain:1993kd}), one can directly compare the
differential cross-section, averaged over the binned recoil energy $T$,
with the data. 
For neutrinos and antineutrinos, the average energies are 
$\braket{E_{\nu_\mu}} = 23.7\,\mathrm{GeV}$ and
$\braket{E_{\bar\nu_\mu}} = 19.1\,\mathrm{GeV}$, respectively.
Since no data correlation from the CHARM-II samples is available, 
we assume all data to follow a gaussian distribution, 
and accordingly define the $\chi^2$ function 
\begin{equation}
\chi^2_\text{CHARM-II} \,=  \,\sum_{i}\left(\frac{\sigma_i - \sigma_{i,
    \text{exp}}}{\Delta \sigma_{i, \text{exp}}} \right)^2\,\text, 
\end{equation}
where $i$ runs over the different bins.
The $\chi^2$ is minimised, and its $1\sigma$ and $2 \sigma$ contours around
the minimum are computed.

\medskip
\paragraph{Data from the TEXONO experiment}

\noindent
The analysis of the TEXONO data~\cite{Deniz:2009mu} is comparatively more
involved than that of CHARM-II. 
Since TEXONO is a reactor experiment, the computation of the binned
event rate requires knowledge of the reactor anti-neutrino flux.
Following the approach of Ref.~\cite{Lindner:2018kjo}, 
the event rate can be computed as
\begin{equation}
R(T_1, T_2) \,= \,
\frac{\rho_e}{T_2 - T_1}\int\phi(E_{\bar\nu})\left[\int_{\bar
    T_1}^{\bar T_2}\frac{d\sigma}{dT} dT \right] d E_{\bar\nu}\,\text, 
\end{equation}
in which $T_{1,2}$ are the bin edges for the electron's recoil energy, 
$\phi(E_{\bar\nu})$ is the neutrino flux, 
$\rho_e$ the electron density of the target material and 
$\bar T_{1,2} = \mathrm{min}(T_{1,2}, T_\text{max})$;
the maximum recoil energy $T_\text{max}$ can be defined as
\begin{equation}
T_\text{max} \,= \,\frac{2E_{\bar\nu}^2}{M + 2 \,E_{\bar\nu}}\,\text.
\end{equation}
The (anti)neutrino flux is given by~\cite{Kopeikin:2004cn}
\begin{equation}
\phi(E_{\bar\nu}) = \frac{1}{4\,\pi\, R^2}\frac{W_{\text{th}}}{\sum_i f_i
  E_{f,i}}\left(\sum_i f_i \rho_i(E_{\bar\nu}) \right)\,\text, 
\end{equation}
in which the sums run over the reactor fuel constituents $i$; for each
of the latter, $f_i$ is the fission rate, 
$E_{f,i}$ the fission energy and $\rho_{i}(E_{\bar\nu})$ the neutrino
spectrum. The remaining intrinsic parameters are 
$W_\text{th}$ - the total thermal energy of the reactor, 
and $R$ which corresponds to the distance between reactor and detector
(details of the reactor and general experimental set-up can be found
in Ref.~\cite{Wong:2006nx}). 
In what concerns the neutrino spectra, and 
depending on the different reactor fuel constituents, we use 
the fit of Ref.~\cite{Mueller:2011nm}, in which 
spectra between $2-8\,\mathrm{MeV}$ are parametrised by 
the exponential of a fifth degree polynomial
\footnote{For completeness, we notice that the lower energy part of
  the spectrum, which is governed by slow neutron capture, 
has been obtained in Ref.~\cite{Kopeikin:1997ve}, 
and is given in the form of numerical results 
for the approximate standard fuel composition of pressurised water
reactors.}.  
(Lower energies are not relevant for our study, since the 
TEXONO data consists of 10 equidistant bins between $3-8\,\mathrm{MeV}$.)

\noindent
We have thus obtained the electron density of the detector material 
$\rho_e$ of the TEXONO experiment by fitting the SM expectation of the 
binned event rate to the SM curve given in Fig.~16 of 
Ref.~\cite{Deniz:2009mu}. Our result is as follows
\begin{equation}
\rho_e \,\simeq \,2.77\times10^{26}\:\mathrm{kg}^{-1}\,\text.
\end{equation}
Finally, and to define the $\chi^2$ function for the TEXONO experiment
data, we again rely on the experimental data Fig.~16 of 
Ref.~\cite{Deniz:2009mu}, leading to
\begin{equation}
\chi^2_\text{TEXONO} \,= \,\sum_{i}\left(\frac{R_i - R_{i,
    \text{exp}}}{\Delta R_{i, \text{exp}}} \right)^2\,\text, 
\end{equation}
where $i$ counts the different bins in the recoil energy.

\begin{figure}
\centering
\includegraphics[width=0.7\textwidth]{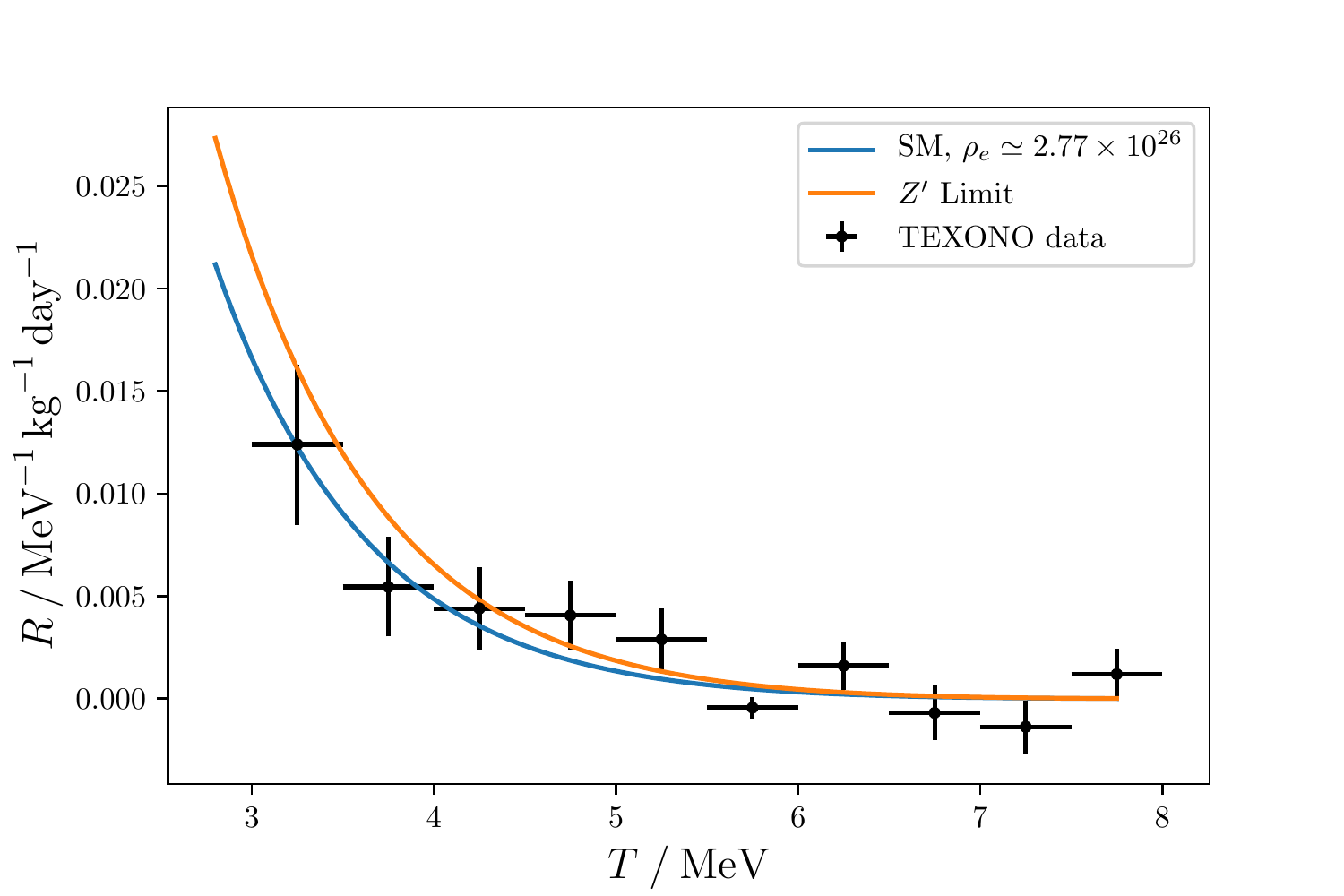}
\caption{Data of the TEXONO experiment (neutrino rate $R$ in units of 
$\mathrm{MeV}^{-1}\,\mathrm{kg}^{-1}\,\mathrm{day}^{-1}$ as a function
  of the binned recoil energy $T$)~\cite{Deniz:2009mu}, to which we superimpose 
our SM and $Z^\prime$ predictions, respectively corresponding to blue and
orange lines.
}
\label{fig:TEXONO_data}
\end{figure}

In Fig.~\ref{fig:TEXONO_data} we display 
the experimental data obtained by TEXONO, together with the (fitted) 
SM curve as well as the $Z^\prime$ prediction. Leading to the 
$Z^\prime$ curve we have taken the minimum couplings to electrons allowed 
by NA64~\cite{Banerjee:2018vgk}, and the maximum values of the 
couplings to neutrinos as derived from the TEXONO data~\cite{Deniz:2009mu}.
The resulting fit is used in the analysis of
Section~\ref{sec:subsec:MajoranaFit}, 
in particular in the results displayed in Fig.~\ref{fig:nu_scat}.

\end{appendix}


\end{document}